\documentclass[preprint,12pt]{elsarticle}
\usepackage[a4paper, left = 2.5cm, right = 2.5cm, top = 2.6cm, bottom = 2.5cm]{geometry}

\usepackage{amssymb}
\usepackage{amsmath}
\usepackage{xurl} % to handle url linebreaks in the bibliography
\usepackage{float} % Allows for figures within multicol and [H] placement
\usepackage{tabularx}
\usepackage{booktabs}
\usepackage{arydshln} % for \cdashline 
\usepackage{bm} % for bold Greek symbols
\usepackage{adjustbox} % for valign=c
\usepackage{subcaption}
\usepackage[bottom]{footmisc} % keep footnotes flushed at bottom of page

\biboptions{sort&compress}

\usepackage[color = lightgrey, colorinlistoftodos, textsize = scriptsize, textwidth = 1.8cm]{todonotes} 
\usepackage{xargs} % Use more than one optional parameter in a new commands
\newcommandx{\todoh}[2][1=]{\todo[disable, #1]{#2}} % Hidden todonotes

\usepackage{datatool}
\DTLsetseparator{,} % Set the separator between the columns.

\DTLloaddb{latex_vars}{latex_vars.dat} % Loads mydata.dat 
\DeclareRobustCommand{\var}[1]{%
  \DTLfetch{latex_vars}{variable}{#1}{value}%
} % DeclareRobustCommand automatically protects arguments (so that \protect\var{} is no longer needed in the captions

\usepackage{caption}
\newcommand{\miniscule}{\fontsize{10}{8}\selectfont}

\usepackage{etoolbox}
\makeatletter
\patchcmd{\pprintMaketitle}
  {\footnotesize\itshape\elsaddress}
  {\miniscule\itshape\elsaddress}{}{}

\patchcmd{\MaketitleBox}
  {\footnotesize\itshape\elsaddress}
  {\miniscule\itshape\elsaddress}{}{}
\makeatother

\newlength{\affilskip}
\makeatletter
\patchcmd{\pprintMaketitle}
  {\elsaddress\par\vskip36pt}
  {\elsaddress\par\vskip\affilskip}{}{}
\patchcmd{\MaketitleBox}
  {\elsaddress\par\vskip36pt}
  {\elsaddress\par\vskip\affilskip}{}{}
\makeatother

\newcommand{\frontmatterfootnotesize}{\scriptsize} % change fontsize here

\makeatletter
\patchcmd{\printFirstPageNotes}
  {\footnotetext{\ifnum\theead=1\relax
      \textit{Email address:\space}\else
      \textit{Email addresses:\space}\fi
     \@elseads}}
  {\footnotetext{\frontmatterfootnotesize
      \ifnum\theead=1\relax
        \textit{Email address:\space}\else
        \textit{Email addresses:\space}\fi
      \@elseads}}
  {}{}

\patchcmd{\printFirstPageNotes}
  {\footnotetext{\textit{URL:\space}\@elsuads}}
  {\footnotetext{\frontmatterfootnotesize\textit{URL:\space}\@elsuads}}
  {}{}

\patchcmd{\cortext}{\footnotetext{#2}}{\footnotetext{\frontmatterfootnotesize #2}}{}{}
\makeatother
\makeatletter
\newcommand{\ackfootnote}[1]{%
  \g@addto@macro\@fnotes{%
    \refstepcounter{fnote}% keep counters in sync
    \global\setcounter{footnote}{\c@fnote}%
    {\let\thefootnote\relax\footnotetext{#1}}% print with no symbol/number
  }%
}
\makeatother

\makeatletter
\DeclareRobustCommand{\abstractfontsize}{\normalsize}

\patchcmd{\keyword}{\normalsize\normalfont}{\abstractfontsize\normalfont}{}{}
\makeatother

\journal{}

\begin{document}
\begin{filecontents*}{latex_vars.dat}
"variable","value"
"new_M_cp_prevalence","0.64"
"new_M_cp_count","5193"
"eligible_rows_count","816,379"
"n_countries","138"
"n_products","1029"
"FDS_years","2011-2023"
"FDS_year1","2011"
"FDS_yearT","2023"
"year_S_E_C_p_scat","2022"
"cor_pci_p_spi_p_2022","0.846"
"cor_pci_p_epi_p_2022","0.738"
"opp_plot_year","2022"
"opp_plot_country_1","Japan"
"opp_plot_country_2","Senegal"
"year_cor_dens","2022"
"mC_1_tpi_ALL_dens_logodds","8.398"
"mC_1_tpi_ALL_dens_logodds_0.1","0.840"
"mC_1_tpi_ALL_dens_oddsratio","2.316"
"mC_2_tpi_ALL_delta_8","0.429"
"mC_2_tpi_ALL_dens_threshold_epi","0.168"
"mC_2_tpi_ALL_dens_threshold_pci","0.281"
"mC_2_tpi_ALL_delta_3","-0.298"
"mC_2_tpi_ALL_delta_6","1.062"
"pct_above_mC_2_tpi_ALL_dens_threshold_pci","20.83"
"dens_cp_P90","0.311"
"mC_2_tpi_HUM_dens_threshold_pci","0.286"
"mC_2_tpi_LML_dens_threshold_pci","0.238"
"mC_2_tpi_ALL_delta_5","-0.072"
"n_countries_above_mC_2_tpi_ALL_dens_threshold_pci","45"
"logistic_threshold_year","2021"
"avg_dens_HUM","0.18"
"avg_dens_LML","0.10"
"mC_2_tpi_HUM_delta_8","0.535"
"mC_2_tpi_LML_delta_8","-0.429"
"avg_dens_cp_topk50_HUM","0.238"
"avg_dens_cp_topk50_LML","0.145"
\end{filecontents*}

%%%% TODO General %%%%
    % todoh: generate a .bib file containing only the references used
    % todoh: remove all doc, readme and manifest files
        % files that can be deleted: README.md, manifest.txt, elsarticle-num.bst, elsarticle-harv.bst

\begin{frontmatter}



\title{\LARGE Economic Complexity Alignment and \\ Sustainable Development}


% Research policy instructions: Affiliations. Add affiliation addresses, referring to where the work was carried out, below the author names. Indicate affiliations using a lower-case superscript letter immediately after the author's name and in front of the corresponding address. Ensure that you provide the full postal address of each affiliation, including the country name and, if available, the email address of each author.

%% Author name
\author[KUL]{{Quinten} {De Wettinck}\corref{corresp}} 
\cortext[corresp]{Corresponding author.}
\ead{quinten.dewettinck@kuleuven.be}

\author[KUL,Stellenbosch]{{Karolien} {De Bruyne}}
\ead{karolien.debruyne@kuleuven.be}

\author[UBC,Stellenbosch]{Wouter Bam}
\ead{wouter.bam@ubc.ca}

\author[Corvinus,Toulouse,Manchester]{César A. Hidalgo}
\ead{cesar.hidalgo@tse-fr.eu}

%% Author affiliations
\affiliation[KUL]{organization={Faculty of Economics and Business, KU Leuven}, % Department and Organization
            addressline={Warmoesberg 26}, 
            city={Brussels},
            postcode={1000}, 
            country={Belgium}
}

\affiliation[UBC]{organization={School of Engineering, University of British Columbia}, 
            addressline={1137 Alumni Ave}, 
            city={Kelowna},
            postcode={V1V 1V7}, 
            state={British Columbia},
            country={Canada}
}

\affiliation[Stellenbosch]{organization={Department of Industrial Engineering, Stellenbosch University}, 
            addressline={145 Banghoek Rd}, 
            city={Stellenbosch},
            postcode={7600}, 
            % state={Western Cape},
            country={South Africa}
}

\affiliation[Corvinus]{organization={Center for Collective Learning, CIAS, Corvinus University of Budapest}, 
            addressline={Közraktár u. 4-6}, 
            city={Budapest},
            postcode={1093}, 
            country={Hungary}
}

\affiliation[Toulouse]{organization={Center for Collective Learning, IAST, Toulouse School of Economics \& Université de Toulouse Capitole}, 
            addressline={1 Esp. de l’Université}, 
            city={Toulouse},
            postcode={31000}, 
            country={France}
}

\affiliation[Manchester]{organization={Alliance Manchester Business School, University of Manchester}, 
            addressline={Booth St W}, 
            city={Manchester},
            postcode={M15 6PB}, 
            country={United Kingdom}
}

% \affiliation[KUL]{organization={}, 
%             country={}
% }    

%% Abstract (max 250 words)
\begin{abstract}
\abstractfontsize
Economic complexity has been linked to sustainability outcomes, such as income inequality and greenhouse gas emissions. Yet, it is unclear whether the pursuit of complex and/or related activities naturally aligns with these outcomes, or whether meeting sustainability goals requires policy interventions that pursue unrelated diversification. Here, we exploit multidimensional social and environmental sustainability indicators to quantify the alignment between a country's closest diversification opportunities and sustainability goals. We find that high- and upper-middle-income countries face significantly better environmentally aligned diversification opportunities than poorer economies. This means that, while richer countries enjoy diversification opportunities that align complexity, relatedness and environmental performance, this alignment is weaker for developing economies. These findings underscore the value of evaluating future diversification trajectories through a multidimensional sustainability framework, and emphasise the strategic relevance of unrelated diversification for less developed economies to foster sustainable development. 

% However, once this linear covariance with complexity is taken out, there is no detectable effect of PSPI on entry

\end{abstract}

%% Graphical abstract
    % You are encouraged to provide a graphical abstract at submission.
    
    % The graphical abstract should summarize the contents of your article in a concise, pictorial form which is designed to capture the attention of a wide readership. A graphical abstract will help draw more attention to your online article and support readers in digesting your research. Some guidelines:
    
    % Submit your graphical abstract as a separate file in the online submission system.
    
    % Ensure the image is a minimum of 531 x 1328 pixels (h x w) or proportionally more and is readable at a size of 5 x 13 cm using a regular screen resolution of 96 dpi.
    
    % Our preferred file types for graphical abstracts are TIFF, EPS, PDF or MS Office files.
    
    % We encourage you to view example graphical abstracts and read about the benefits of including them.

% \includegraphics[width = \textwidth]{WP1_graphical_abstract.pdf}

%% Research highlights
% Must be provided as a Word document
% Max 85 characters (spaces included)
% See research_highlights.docx

%% Keywords

% Acknowledgements
\ackfootnote{\frontmatterfootnotesize {\frontmatterfootnotesize \textit{Acknowledgements:}} We acknowledge the support of the European Union LearnData, GA no. 101086712 a.k.a. 101086712-LearnDataHORIZON-WIDERA-2022-TALENTS-01 (\url{https://cordis.europa.eu/project/id/10IAST}), funding from the French National Research Agency (ANR) under grant ANR-17-EURE0010 (Investissements d’Avenir program), and the European Lighthouse of AI for Sustainability grant number 101120237-HORIZON-CL4-2022-HUMAN-02. We also acknowledge the support of the Natural Sciences and Engineering Research Council of Canada (NSERC), [funding reference number RGPIN-2024-04381].
Cette recherche a été financée par le Conseil de recherches en sciences naturelles et en génie du Canada (CRSNG), [numéro de référence RGPIN-2024-04381]. Additionally, we are grateful to Paul O. Bekambo for valuable discussions that enriched this work.}

% \ackfootnote{\tiny {\frontmatterfootnotesize \textit{Declaration of interest:}} The authors declare no competing financial or personal interests.}

\end{frontmatter}



\section{Introduction}

The sustainability of economies is contingent on preserving a healthy natural environment, building a strong civic society, and generating the means required to secure, educate, and support a flourishing population \citep{Perrings1994, Nogueira.etal2023}. This notion is embodied in calls for a triple bottom line (TBL), encompassing economic, social, and environmental objectives. As highlighted by \citet{Boleti.etal2021}, performance across these three TBL dimensions is largely impacted by the goods and services an economy produces and exports. 

The economic complexity literature provides a powerful framework to analyse economic diversification, by modelling it as a capability-constrained, path-dependent process that can be inferred from observed specialisation networks \citep{Hidalgo.etal2007, Hidalgo.Hausmann2009, Hausmann.Hidalgo2011, Hidalgo2021, Pinheiro.etal2022}. While this field has established a strong empirical link between measures of economic complexity and many country-level sustainability outcomes \citep{Caldarola.etal2024, Hidalgo2021}, a precise understanding of how product-level complexity and sustainability jointly impact diversification dynamics is still lacking \citep{Montiel-Hernandez.etal2024, Ferraz.etal2021, Fraccascia.etal2018}. More specifically, it remains unclear whether the pursuit of complexity inherently aligns with sustainability gains across all stages of economic development, or whether some countries face unavoidable trade-offs between these dimensions. 

To address these shortcomings, we model the trade-offs between product complexity, social, and environmental sustainability. We achieve this by combining multidimensional social and environmental indicators to investigate the alignment or correspondence between the complexity and sustainability of the most related specialisation opportunities available to countries. That is, we explore whether the most likely diversification paths faced by countries are also the most promising paths for social and environmental sustainability, or whether there are trade-offs among these different directions. 

Using a country-product export specialisation network, we first compute measures of relatedness and economic complexity, and exploit country-level sustainability scores to construct indicators of product-level social and environmental sustainability. These metrics are then used to investigate related diversification opportunities, and quantify the degree of alignment between the complexity, the social and the environmental performance of these products. 

We find that the aggregate alignment between the three dimensions is generally positive, though important differences exist between income groups, especially for environmental performance. Additionally, we model diversification patterns through product entries, and find that countries are more likely to enter related complex products over related socially or environmentally sustainable ones. Notably, for richer countries, both product complexity and environmental performance raise the odds of entry for related products, whereas diversification in poorer economies appears to be driven predominantly by complexity.

The remainder of this paper is organised as follows. In Section~\ref{complexity_and_sustainable_development}, we introduce the main concepts from the traditional economic complexity framework and summarise the existing literature on the connection between complexity and sustainable development. Section~\ref{methodology} outlines the data sources and methods used in the analysis, and Section~\ref{results} presents the results. Finally, Section~\ref{discussion} provides the interpretation of these findings, which includes drawing broader conclusions about the role of opportunity alignment in achieving equitable and environmentally responsible economic development.

\section{Economic Complexity and Sustainable Diversification}\label{complexity_and_sustainable_development}

Economic complexity has become a standard methodology to examine the diversification opportunities available to individual countries, regions, and cities \citep{Hidalgo.etal2007, Hidalgo.Hausmann2009, Neffke.etal2011, Kogler.etal2013, Muneepeerakul.etal2013, Neffke.Henning2013, Boschma.etal2013, Boschma.etal2015, Poncet.DeWaldemar2015, Rigby2015, Guevara.etal2016, Borggren.etal2016, Chen.etal2017, Jara-Figueroa.etal2018, Pinheiro.etal2018, Alabdulkareem.etal2018, Farinha.etal2019, Hartmann.etal2019, Chinazzi.etal2019, Cicerone.etal2020, Juhasz.etal2021, Hartmann.etal2021, Bam.etal2021, Pinheiro.etal2022, Bam.DeBruyne2022, DeBruyne.etal2023, Stephany.Teutloff2024, Pinheiro2024, Timbe.etal2024, Pinheiro.etal2025}. Identifying viable sustainable diversification options for a country entails finding a set of products which are both \emph{attainable} for the country to produce and export, as well as \textit{desirable} in terms of economic, social and environmental sustainability objectives \citep{Bam.DeBruyne2022}. To find such products, economic complexity scholars usually use two concepts: measures of relatedness and measures of product-level performance (such as the product complexity index \citep{Hidalgo.Hausmann2009}, product Gini index \citep{Hartmann.etal2017}, or product emission intensity index \citep{Romero.Gramkow2021}). 

Relatedness metrics assess how likely a country is to diversify into a particular product, given its current pattern of specialisation. Economic complexity captures the technological sophistication of the capabilities present within countries. Countries that export products that are complex for their level of income tend to grow faster, making diversification into such products desirable from an economic point of view \citep{Hidalgo.Hausmann2009}. Similarly, product sustainability indicators reflect the association between exporting products and country-level sustainability outcomes \citep{Hartmann.etal2017, Lapatinas.etal2019, Mealy.Teytelboym2022}. 

\subsection{Observed Networks and Latent Capabilities}

In the economic complexity literature, complexity and relatedness measures are extracted from a specialisation network connecting economies (e.g. countries, regions, cities, ...) to the activities they specialise in (e.g. exporting products, obtaining patents, publishing research, ...). In the context of trade diversification, export data is commonly used to connect countries to the products they specialise in. The main assumption underlying this approach is that countries are able to specialise in the export of a product if they possess the required capabilities. These capabilities or `productive factors' serve as a hidden layer connecting the nodes in the observed bipartite country-product network \citep{Hidalgo.Stojkoski2025}. These capabilities are often referred to as `genotypic', meaning they are not directly observable but should instead be inferred using information of the `phenotypic', observable specialisation network \citep{Balland.etal2022, Schetter.etal2024}. Generally, capabilities include all endowments that make the production and export of tradable goods possible \citep{Cristelli.etal2013, Cristelli.etal2017}, such as regulatory frameworks, infrastructure, norms, social properties, and other nontradable activities \citep{Hausmann.Hidalgo2011}. In contrast to traditional economic frameworks, this approach notably remains agnostic about the exact nature of these capabilities, thereby removing the need for stringent assumptions about their properties \citep{Hidalgo2021}. 

Since it is assumed that the country-capability-product network induces the observed country-product export network, the latter one can be used to quantify three types of measures that are required to identify sustainable diversification paths: relatedness, complexity and sustainability measures. Section~\ref{methodology} expands on how exactly these metrics are computed using international trade data. 

\subsection{Relatedness and Economic Complexity}

Extensive prior research has demonstrated that diversification is a highly path-dependent process \citep{Pinheiro.etal2022, Hausmann.Hidalgo2011, Hidalgo.etal2007, Neffke.etal2011, Guevara.etal2016, Chinazzi.etal2019, Kogler.etal2013, Hidalgo2021, Hidalgo.etal2018, Boschma.Capone2015, Zhu.etal2017}. These path dependencies are known as the \textit{principle of relatedness}, which captures the notion that a country is more likely to specialise in products that require similar capabilities to the ones used by products it is already specialised in \citep{Hidalgo.etal2018, Hidalgo2021, Tacchella.etal2023}. However, identifying products that serve as attainable diversification opportunities using individual product characteristic data is infeasible, since it would require an overwhelming amount of information that is typically unavailable at such a granular level \citep{Hausmann.Hidalgo2013}. Hence, the country-product specialisation network is used to compute relatedness metrics that help identify attainable diversification opportunities for individual countries. First, the topology of the network is used to determine the similarity or \textit{proximity} between pairs of products. Thereafter, these product proximities can be used in combination with the network structure to quantify the \textit{relatedness densities} between countries and products, providing a measure of how likely a country is to specialise in that product. 

Next to that, the economic complexity of a country refers to the level of sophistication of the capabilities expressed in the products it specialises in. Similarly, the complexity of a product captures the sophistication of the capabilities that are required to produce and export that product. This recursive characterisation implies that the complexity of a country is dependent on the complexity of the products it exports, and the complexity of a product is a function of the complexity of the countries it is exported by \citep{Hidalgo.Hausmann2009}. 

\subsection{Complexity and Sustainability}

A country's specialisation pattern is predictive of both economic growth and sustainable development outcomes \citep{Domini2019, Hidalgo.Hausmann2009, Koch2021, LoTurco.Maggioni2022, Ourens2013, Stojkoski.etal2016, Cristelli.etal2017, Felipe.etal2012, Stojkoski.etal2023, Zhu.Li2017, Hidalgo2021}. For instance, a large body of literature finds that economic complexity is related to lower greenhouse gas emissions \citep{Romero.Gramkow2021, Dogan.etal2021, Safi.etal2023, Saqib.Dinca2023, Stojkoski.etal2023}, reduced emission intensity \citep{Romero.Gramkow2021, Payne.etal2023}, more biodiversity \citep{BalsalobreLorente.etal2023}, and increased overall environmental performance \citep{Boleti.etal2021}.

However, several empirical analyses on this relationship provide mixed results, hinting at a nonlinear link between complexity and environmental performance \citep{Caldarola.etal2024}. For instance, some studies find that complexity is related to \textit{increased} ecological footprint \citep{Rafique.etal2022}, and worse air quality \citep{Boleti.etal2021}. This apparent nonlinearity is reminiscent of the dynamics described by the environmental Kuznets curve, which postulates an inverted U-shaped relationship between economic development and environmental degradation \citep{Grossman.Krueger1991, Grossman.Krueger1995}. Kuznets type of relationships have been observed between economic complexity and ecological footprint \citep{Kilic.etal2024}, as well as carbon emissions \citep{Neagu2019}. This concave relationship between complexity and environmental performance is rooted in the observation that more complex economies tend to shift their focus towards more knowledge-intensive goods, which are often greener \citep{Caldarola.etal2024}. This aligns with the findings of \citet{Dogan.etal2022}, who observe that complexity appears to drive energy-based demand in the short term, but decreases it in the long run. 

Moreover, complexity has been found to be associated with social outcomes, such as reduced income inequality \citep{Hartmann.etal2017, Fawaz.Rahnama-Moghadamm2019, Stojkoski.etal2023}, more LGBTQ+ inclusion \citep{Vu2020a, Vu2022a}, improved national health \citep{Vu2020}, increasing secondary and higher education \citep{Zhu.Li2017}, less gender inequality \citep{BenSaad.Assoumou-Ella2019, Barza.etal2020}, reduced unemployment \citep{Adam.etal2023}, and better institutional quality \citep{Nguyen.etal2023, Vu2022}. Analogously to the environmental dimension, some studies report evidence for a nonlinear connection between complexity and income inequality \citep{Morais.etal2018, Sbardella.etal2017, Nguyen.etal2023, Ghosh.etal2023}, human development \citep{LeCaous.Huarng2020}, and obesity \citep{Djeunankan.etal2025}.

However, to obtain a more nuanced view of the complexity-sustainability connection in the context of trade diversification, scholars utilise the structure of the country-product specialisation network to construct measures of product-level sustainability. To avoid issues of model unidentifiability, the conventional approach to estimate the association between product exports and country-level sustainability is to express the performance of a product as the weighted average of the scores of the countries that specialise in that product (see equation~(\ref{eq_hartmann}) on p\pageref{eq_hartmann}), which was introduced with measures such as PRODY \citep{Hausmann.etal2007}, Product Gini Index \citep{Hartmann.etal2017}, and Product Environmental Performance Index \citep{Lapatinas.etal2019}. 

The major advantage of this approach is that there is no need for a predetermined list of `green' or `social' products to be relied upon, since continuous performance scores can be assigned to each product using information contained in the country-product specialisation network \citep{Fraccascia.etal2018}. Nevertheless, many of the existing metrics using this approach consider only rather `narrow' aspects of sustainability. For instance, existing metrics linking product exports to social sustainability have focused on either income inequality \citep{Hartmann.etal2017} or unemployment \citep{Adam.etal2023}, not accounting for the many other factors comprising social sustainability such as, among others, poverty, human rights, freedom, and safety. Similarly, studies investigating the connection between products and environmental sustainability often focus predominantly on greenhouse gas emissions \citep{Romero.Gramkow2021}, ignoring other environmental elements such as biodiversity, water and plastic pollution, energy, or resource usage.

For these reasons, this study aims to contribute to this literature by integrating economic complexity and relatedness measures with comprehensive social and environmental sustainability indicators to study export diversification opportunities and their level of sustainability alignment. 

\section{Data and Methods}\label{methodology}

The following sections outline our methodological approach. Section~\ref{data} introduces the data sources, followed by Section~\ref{specialisation_network}, which details the construction of the country-product specialisation network. Sections~\ref{relatedness}, \ref{economic_complexity}, \ref{p_lvl_metrics}, and \ref{methods_opportunities} expand on the calculation of the measures of relatedness, economic complexity, product-level sustainability, and opportunity alignment, respectively. Finally, Section~\ref{diversification_dynamics} delineates the empirical specification strategy for modelling the nonlinear connections between export diversification, complexity, and sustainability. 

\subsection{Data}\label{data}

The country-product specialisation network is constructed using the Base pour l'Analyse du Commerce International (BACI) trade data provided by the Centre d'Etudes Prospectives et d'Informations Internationales (CEPII), containing bilateral trade flows at the 4-digit level of the 2007 edition of the Harmonised System (HS07) nomenclature \citep{Gaulier.Zignago2010}. Additionally, this dataset is extended with the Social Progress Index (SPI) and Environmental Performance Index (EPI) to capture the social and environmental sustainability of countries. The SPI, provided by the \citet{SocialProgressImperative2024}, aggregates data of 57 subindicators of social sustainability into an index representing ``the capacity of a society to meet the basic human needs of its citizens'', and covers subjects such as education, freedom, human rights, inclusiveness, health, housing, access, and safety \citep{EuropeanCommission2023a, Stern.etal2024}. Similarly, the EPI combines 58 individual indicators of environmental performance, grouped into the categories climate change, environmental health and ecosystem vitality \citep{Block.etal2024}. Since the definition and weighting scheme of the EPI have been revised between editions, the values of the biannually released index are not comparable over time. Hence, we use the 2024 indicator weights to compute the EPI from the indicator time series that is made available by The \citet{YaleCenterforEnvironmentalLaw&Policy2024}. Further details on the specific subcomponents of the SPI and EPI are provided later in Section~\ref{p_lvl_metrics}. Lastly, country-level covariates are collected from the World Development Indicators (WDI) of \citet{TheWorldBank2025}. 

Consistent with prior literature \citep{Hartmann.etal2017, Pinheiro.etal2022, Stojkoski.etal2023, Jun.etal2020, Hartmann.etal2021}, we exclude small economies by retaining only countries with total exports exceeding $1$ billion USD, and a population larger than $1$ million in 2011. Additionally, we remove smaller products with global exports lower than $500$ million USD in 2011. This minimises distortions arising from the volatile specialisation patterns of small countries or products \citep{Stojkoski.etal2023}, and results in a final sample of \var{n_countries} countries and \var{n_products} products from \var{FDS_year1} to \var{FDS_yearT}. 

\subsection{Specialisation Network}\label{specialisation_network}

We aggregate the BACI data into a bipartite network connecting countries $c \in \{1, \dots, C\}$ to the products $p \in \{1, \dots, P\}$ they specialise in. Given the vast scale differences between the raw export values across countries (e.g. China versus Uganda), we normalise trade flows to make the units of observation comparable. We follow the conventional normalisation procedure for export data by using the \emph{revealed comparative advantage} (RCA) measure introduced by \citet{Balassa1965}, which is the ratio of a country's share of exports in product $p$, to the share of $p$ in world trade. Formally, the RCA of country $c$ in product $p$ is given by\footnote{In this literature it is conventional to use Einstein notation, in which missing indices represent summed variables, i.e. $A_{i} = \sum_j A_{ij}$ for any matrix $\mathbf{A} = [A_{ij}]$ \citep{Hidalgo2021}.}
\begin{equation}\label{eq_RCA}
    \mathit{RCA}_{cp} = \frac{X_{cp} / \sum_{p'} X_{cp'}}{\sum_{c'} X_{c'p} / \sum_{c'p'} X_{c'p'}}
\end{equation}
where $X_{cp}$ are the exports of country $c$ in product $p$ (often expressed in USD). Thereafter, we binarise the RCAs by constructing a $C \times P$ dimensional binary adjacency matrix $\mathbf{M}$, whose elements are defined as
\begin{equation}\label{eq_M_cp}
    M_{cp} = \begin{cases} 1 & \text{if} \ \mathit{RCA}_{cp} \geq 1 \\ 0 & \text{if} \ \mathit{RCA}_{cp} < 1 \end{cases}
\end{equation}
which encodes information about the topology of the specialisation network by indicating which products are exported with comparative advantage by each country. This binarisation further denoises the matrix, retaining only the signal identifying which products countries are specialised in. 

\subsection{Relatedness}\label{relatedness}

To quantify the likelihood of a country entering a product it is currently not specialised in (i.e. $\mathit{RCA}_{cp} < 1$), we estimate relatedness density using an algorithm that is equivalent to collaborative filtering methods in recommender systems \citep{Hidalgo2021}. First, we compute the similarity or \emph{proximity} between pairs of products based on their co-occurrences in the specialisation network $\mathbf{M}$. The proximity between products $p$ and $p'$ is based on their normalised co-occurrences in the $M_{cp}$ export network, which is formally given by
\begin{equation}\label{eq_proximity}
    \phi_{pp'} = \frac{1}{\max(M_p, M_{p'})} \sum_c M_{cp} M_{cp'}
\end{equation}
where $M_p = \sum_c M_{cp}$ is the \emph{ubiquity} of product $p$ (similarly $M_c = \sum_p M_{cp}$ is the \emph{diversity} of country $c$) \citep{Hidalgo.etal2007}. Hence, $\phi_{pp^\prime} \in [0, 1]$ represents the minimum conditional probability that a country $c$ would possess a comparative advantage in both $p$ and $p^\prime$, given that it exports at least one of the two. 

We then use these proximities to quantify the \emph{relatedness density} between countries and products, which indicates how likely it is that a country will be able to obtain an $\mathit{RCA}_{cp} \ge 1$ in the future, given its current set of capabilities. Formally, the relatedness density between country $c$ and product $p$ is given by 
\begin{equation}\label{eq_density}
    \omega_{cp} = \frac{1}{\sum_{p^\prime} \phi_{pp^\prime}}\sum_{p^\prime} M_{cp^\prime} \phi_{pp^\prime}
\end{equation}
which will serve as our measure for the attainability of future diversification opportunities.

\subsection{Economic Complexity}\label{economic_complexity}

To capture the technological sophistication of countries and products, we use the economic complexity measure proposed by \citet{Hidalgo.Hausmann2009}. In this approach, the level of complexity of both countries and products is quantified using an algorithm akin to matrix factorisation techniques such as singular value decomposition, in which the structure of the specialisation matrix is expressed in lower-dimensional vectors \citep{Hidalgo2021}. In particular, we employ the original method introduced by \citet{Hidalgo.Hausmann2009}, which recursively links country complexity $K_c$ and product complexity $K_p$ through the system 
\begin{equation}
    \left \{ 
    \begin{aligned}
        K_c & = \frac{1}{M_c} \sum_p M_{cp} K_p \\
        K_p & = \frac{1}{M_p} \sum_c M_{cp} K_c.
    \end{aligned}
    \right . 
\end{equation}
Since these arithmetic means are linear equations, they imply the self-consistent equations
\begin{equation}
    \left \{
    \begin{aligned}
        K_c & = \tilde{M}_{cc'} K_{c'} \\
        K_p & = \tilde{M}_{pp'} K_{p'}
    \end{aligned}
    \right . 
\end{equation}
where the elements of the square matrices are defined as 
\begin{equation}
    \tilde{M}_{cc'} = \sum_p \frac{M_{cp} M_{c'p}}{M_c M_p} \quad \text{and} \quad \tilde{M}_{pp'} = \sum_c \frac{M_{cp} M_{cp'}}{M_c M_p} 
\end{equation}
\citep{Hidalgo2021, Mealy.etal2019, Caldarelli.etal2012, Cristelli.etal2013}. The eigenvectors of this system associated with the second largest eigenvalues capture the largest amount of variance in the system \citep{Hausmann.Hidalgo2013}, and are therefore used to quantify the complexity of countries and products\footnote{The eigenvectors with the \emph{second} largest eigenvalues are used since $[\tilde{M}_{cc'}]$ and $[\tilde{M}_{pp'}]$ are row-stochastic matrices with a first eigenvector of $1$s, hence carrying trivial information \citep{Hidalgo2021, Sciarra.etal2020, Caldarelli.etal2012}.}. Because of the relative nature of the complexity metrics, these eigenvectors ($\vec{K}$ for $[\tilde{M}_{cc'}]$ and $\vec{L}$ for $[\tilde{M}_{pp'}]$) are standardised using a $Z$-transformation, resulting in the following expressions for the economic complexity index and product complexity index: 
\begin{equation}\label{eq_eci_pci}
    \left \{
    \begin{aligned}
        \mathit{ECI}_c & = \frac{\vec{K} - \mu(\vec{K})}{\sigma(\vec{K})} \\
        \mathit{PCI}_p & = \frac{\vec{L} - \mu(\vec{L})}{\sigma(\vec{L})}
    \end{aligned}
    \right . 
\end{equation}
where $\mu(\cdot)$ and $\sigma(\cdot)$ are the sample mean and standard deviation of the eigenvectors, respectively \citep{Hidalgo.Hausmann2009, Hidalgo2021}. 

% Motivation for ECI: 
    % for the sake of cross-study comparability? 
    % orthogonality with diversity 

\subsection{Product-Level Sustainability Measures}\label{p_lvl_metrics}

In contrast to the economic complexity measures, product social and environmental sustainability metrics cannot be derived solely from the country-product trade network. Instead, they rely on external data capturing country-level performance on these dimensions. As discussed in Section~\ref{complexity_and_sustainable_development}, the conventional way to convert country to product scores is to express the product score $T_p$ as the weighted average of the scores of the countries that export $p$ with an $\mathit{RCA}_{cp} \ge 1$. Formally, this is expressed as 
\begin{equation}\label{eq_hartmann}
    T_p = \frac{1}{\sum_c M_{cp} s_{cp}} \sum_c M_{cp} s_{cp} T_c
\end{equation}
% $\tilde{T}_p = \frac{1}{\sum_c M_{cp} s_{cp}} \sum_c M_{cp} s_{cp} Z(T_c)$ where $Z(X) = \frac{X - \bar{X}}{SD(X)}$, and then $T_p = Z(T_p)$ (cfr. double standardisation 10-10-2024)
where $T_c$ is the performance score of country $c$, $s_{cp} = X_{cp} / \sum_{p'}X_{cp'}$ represents the share of product $p$ in the export basket of country $c$, and $M_{cp}$ denotes the elements of the binarised specialisation matrix, which `activates' the export shares to only consider countries that have an $\mathit{RCA}_{cp} \ge 1$ \citep{Hartmann.etal2017}. $T_p$ is therefore an estimate of the expected value of the performance measure $T_c$ among heavy exporters of $p$. 

Here, we consider performance measures collected from the Social Progress Index (SPI) and Environmental Performance Index (EPI). This technique has been used by several studies using the Gini coefficient \citep{Hartmann.etal2017}, unemployment \citep{Adam.etal2023}, emission intensity \citep{Romero.Gramkow2021}, air pollution \citep{Lapatinas.etal2019}, and the EPI \citep{Lapatinas.etal2019} itself. Tables~\ref{tab_SPI} and \ref{tab_EPI} provide an overview of all the subdimensions contained in the SPI and EPI, respectively.

\begin{table}[H]
  \centering
  \caption{Subcomponents of the Social Progress and Environmental Performance Indices}
  \begin{subtable}[t]{.48\textwidth}
    \caption{Social Progress Index}
    \label{tab_SPI}
    % ↓ ↓ smaller font + tighter columns only inside this group
    {\scriptsize\setlength{\tabcolsep}{3pt}
    \begin{tabularx}{\linewidth}{@{}>{\raggedright\arraybackslash}p{.38\linewidth} X@{}}
      \toprule
      \multicolumn{1}{c}{Category} & \multicolumn{1}{c}{Component} \\
      \midrule
      Basic Needs & 
        \adjustbox{valign=c}{\begin{tabular}{@{}l@{}}Nutrition and Medical Care\\Water and Sanitation\\Housing\\Safety\end{tabular}} \\ \cdashline{1-2}\addlinespace[2pt]
      Foundations of Wellbeing & 
        \adjustbox{valign=c}{\begin{tabular}{@{}l@{}}Basic Education\\Information and \\ \hspace{1em} Communications \\ Health\\Environmental Quality\end{tabular}} \\ \cdashline{1-2}\addlinespace[2pt]
      Opportunity & 
        \adjustbox{valign=c}{\begin{tabular}{@{}l@{}}Rights and Voice\\Freedom and Choice\\Inclusive Society\\Advanced Education\end{tabular}} \\
      \bottomrule
    \end{tabularx}}
    \vspace{0.4em}
    \parbox{\linewidth}{\vspace{0.2em}\scriptsize\textit{Note:} Full indicator list in \citet{Stern.etal2024}.}
  \end{subtable}
  \hfill
  \begin{subtable}[t]{.48\textwidth}
    \caption{Environmental Performance Index}
    \label{tab_EPI}
    {\scriptsize\setlength{\tabcolsep}{3pt}
    \begin{tabularx}{\linewidth}{@{}>{\raggedright\arraybackslash}p{.38\linewidth} X@{}}
      \toprule
      \multicolumn{1}{c}{Category} & \multicolumn{1}{c}{Component} \\
      \midrule
      Ecosystem Vitality & 
        \adjustbox{valign=c}{\begin{tabular}{@{}l@{}}Biodiversity \& Habitat\\Forests\\Fisheries\\Air Pollution\\Agriculture\\Water Resources\end{tabular}} \\ \cdashline{1-2}\addlinespace[2pt]
      Environmental Health & 
        \adjustbox{valign=c}{\begin{tabular}{@{}l@{}}Air Quality\\Sanitation \& Drinking \\ \hspace{1em} Water\\Heavy Metals\\Waste Management\end{tabular}} \\ \cdashline{1-2}\addlinespace[2pt]
      Climate Change & 
        \adjustbox{valign=c}{\begin{tabular}{@{}l@{}}Climate Change Mitigation\\ \\\end{tabular}} \\
      \bottomrule
    \end{tabularx}}
    \vspace{0.4em}
    \parbox{\linewidth}{\vspace{0.2em}\scriptsize\textit{Note:} Full indicator list in \citet{Block.etal2024}.}
  \end{subtable}
\end{table}

Applying equation~(\ref{eq_hartmann}) to $\mathit{SPI}_c$ and $\mathit{EPI}_c$ yields two metrics for product-level social and environmental sustainability, referred to as the Product Social Progress Index ($\mathit{PSPI}_p$) and Product Environmental Performance Index ($\mathit{PEPI}_p$) \citep{Lapatinas.etal2019}. As with the $\mathit{ECI}_c$, the $\mathit{SPI}_c$ and $\mathit{EPI}_c$ are first standardised using a $Z$-transform to facilitate comparison across indices. 

\subsection{Opportunity Alignment}\label{methods_opportunities} 

We then use relatedness density, complexity, and the two product sustainability measures $\mathit{PSPI}_p$ and $\mathit{PEPI}_p$ to study the alignment between the complexity and sustainability of the opportunities available to countries. To estimate this level of opportunity alignment, we focus on the closest diversification opportunities available to each country, which are contained in the set
\begin{equation}
    \Omega_{ct} \equiv \{p^\prime \mid \omega_{cp^\prime t} \in \mathrm{top}_k ({\omega_{cpt} \mid M_{cpt} = 0})\}_{ct}
\end{equation} 
where $\mathrm{top}_k(\cdot)$ returns the $k$ largest relatedness density values for products in which $c$ does not have a revealed comparative advantage at time $t$. This idea of an option set has been used by \citet{Pinheiro.etal2022} and \citet{Timbe.etal2024}. Here, we apply this concept but restricted to a subset of the $k$ closest products to each country, which contains only the most likely diversification opportunities given a country's current specialisation pattern.

\begin{figure}[H]
    \centering
    \includegraphics[width=1\linewidth]{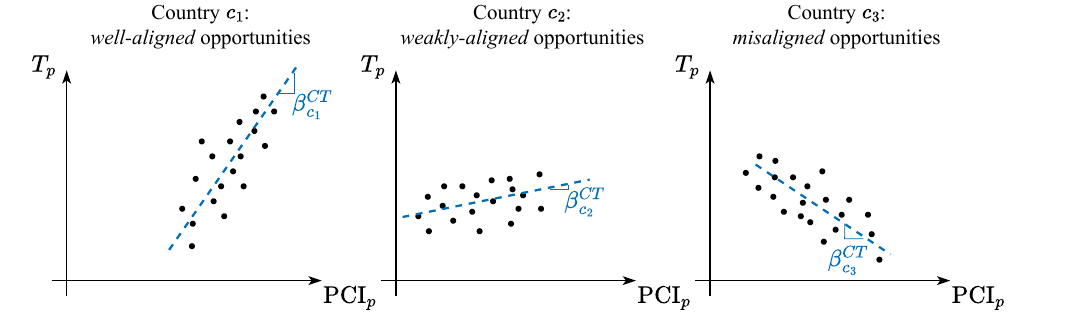}
    \caption{Conceptual Diagram of the Complexity-Sustainability Alignment of the Closest Opportunities to Countries}
    \label{fig_slope_diagram}
\end{figure}

We then evaluate the sustainability alignment of individual diversification trajectories by examining the strength of the relationship between the complexity and social or environmental scores of these closest opportunities. First, we plot these opportunities for each country in a complexity-sustainability plane, which is informative of the direction each country is expected to diversify into (see Fig.~\ref{fig_slope_diagram}). Thereafter, we quantify the alignment between the complexity and sustainability of these opportunities using the estimated slope coefficient $\hat{\beta}^{CT}_{ct}$ of a regression predicting the product sustainability indicator $T_{pt}$ (which is either $\mathit{PSPI}_{pt}$ or $\mathit{PEPI}_{pt}$) from $\mathit{PCI}_{pt}$ for all products in the closest opportunity set $\Omega_{ct}$.

Since these are simple linear regression models, these alignment measures can be explicitly written as 
% \begin{equation}\label{eq_slope}
%     \hat{\beta}^{CT}_c \equiv \frac{\sum_{p \in \Omega_c}(C_p - \bar{C}_p)(T_p - \bar{T}_p)}{\sum_{p \in \Omega_c} (C_p - \bar{C}_p)^2}. 
% \end{equation} 
% \begin{equation}\label{eq_slope}
%     \hat{\beta}^{CT}_{ct} \equiv \frac{\sum\limits_{p \in \Omega_{ct}}(C_{pt} - \bar{C}_{pt})(T_{pt} - \bar{T}_{pt})}{\sum\limits_{p \in \Omega_{ct}} (C_{pt} - \bar{C}_{pt})^2}
% \end{equation} 
\begin{equation}\label{eq_slope}
    \hat{\beta}^{CT}_{ct} = \frac{\sum\limits_{p \in \Omega_{ct}}(\mathit{PCI}_{pt} - \bar{\mathit{PCI}}_{pt})(T_{pt} - \bar{T}_{pt})}{\sum\limits_{p \in \Omega_{ct}} (\mathit{PCI}_{pt} - \bar{\mathit{PCI}}_{pt})^2}
\end{equation} 
where $\bar{\mathit{PCI}}_{pt}$ and $\bar{T}_{pt}$ are the sample means of product complexity and product sustainability for all $p \in \Omega_{ct}$. A large and positive value for $\hat{\beta}^{CT}_{ct}$ means that the country is likely to enter products that perform well in both complexity and sustainability (e.g. country $c_1$ in Fig.~\ref{fig_slope_diagram}). A weakly positive slope indicates that while increased complexity correlates with higher sustainability, these gains remain comparatively weaker (e.g. country $c_2$). A negative value means that the complexity and sustainability of its nearest opportunities are misaligned, since they perform well in only one of the two metrics (e.g. country $c_3$). In that case, a country's most likely diversification path is incompatible with increasing the value of both dimensions, implying hard trade-offs.

The reason for only considering the top $k$ most related products for each country is to introduce country-level variation in these complexity-social ($\hat{\beta}^{CS}_{ct}$) and complexity-environmental ($\hat{\beta}^{CE}_{ct}$) alignment measures. After all, using \textit{all} unattained products would give estimates that are approximately equal to the global slope between $\mathit{PCI}_{pt}$ and $T_{pt}$. A threshold of $k = 50$ is adopted for the results presented below. However, the findings remain consistent for alternative specifications with $k = 30$ and $k = 100$, with larger values of $k$ leading to greater convergence across countries as this diminishes the influence of relatedness heterogeneity on the alignment metrics. 

\subsection{Product Entry Dynamics}\label{diversification_dynamics}

We define an entry or successful diversification as a country developing a \textit{sustained} $\mathit{RCA}_{cp} \ge 1$ in a product $p$. To mitigate the risk of false positives caused by temporary fluctuations, we use both a backward and forward condition \citep{Pinheiro.etal2018}. The backward condition ($B_{cpt}$) mandates that RCA was smaller than $1$ for $m = 3$ consecutive years before the entry, while the forward condition ($F_{cpt}$) requires the country to sustain an RCA $\ge 1$ for the $3$ subsequent years. These conditions make the entry measure more stable, enhancing the robustness of the results. Missing values of $M_{cpt}$ are excluded. Formally, this binary outcome is defined as
% \begin{equation} \label{eq_new_M_cp}
%     \Delta M_{cpt} \equiv \mathbb{I} \Bigg( \underbrace{ \vphantom{\bigwedge\limits_{t^\prime = 0}^{m - 1}} \bigwedge\limits_{t^\prime = 1}^{m} (M_{cpt - t^\prime} = 0) }_{B_{cpt}}
%     \land 
%     \underbrace{ \vphantom{\bigwedge\limits_{t^\prime = 1}^{m}} \bigwedge\limits_{t^\prime = 0}^{m - 1} (M_{cpt + t^\prime} = 1) }_{F_{cpt}} \Bigg)
% \end{equation}
\vspace{0.8cm}
\begin{equation}
    \Delta M_{cpt} \equiv
    \begin{cases}
    1 \text{ if } \quad
    \smash[t]{\overbrace{\vphantom{\bigwedge\limits_{t' = 0}^{m-1}(M_{cpt+t'}=1)}
\,\bigwedge\limits_{t' = 1}^{m}(M_{cpt - t'} = 0)}^{\text{Backward cond. } B_{cpt}}}
    \;\land\;
    \smash[t]{\overbrace{\vphantom{\bigwedge\limits_{t' = 0}^{m-1}(M_{cpt+t'}=1)}
    \,\bigwedge\limits_{t' = 0}^{m-1}(M_{cpt + t'} = 1)}^{\text{Forward cond. } F_{cpt}}}
    \\
    0 \text{ else}
    \end{cases}
\end{equation}
    % \Delta M_{cpt} can also be expressed as \Delta M_{cpt} = M_{cpt} - M_{cpt-1}
which simply checks whether both the backward $B_{cpt}$ and forward condition $F_{cpt}$ are simultaneously satisfied. We then use this entry measure to examine the relative effects of product complexity, social and environmental performance on diversification dynamics. More specifically, we test whether countries tend to move in systematically distinct directions within the complexity-sustainability space presented in Fig.~\ref{fig_slope_diagram}. 

To assess the relative importance of complexity versus sustainability on export diversification, we empirically model the probability of product entry as a function of relatedness density, log RCA, product complexity, and product sustainability, all lagged by one year, as well as the two-way interactions between relatedness and the three product performance measures. These interaction terms with relatedness reflect the intrinsic nonlinearities driving the diversification process by incorporating heterogeneity in countries' underlying capability structures. Additionally, to avoid multicollinearity issues due to strong correlations between product complexity and sustainability, the product sustainability indicators are orthogonalised with respect to PCI. Specifically, each sustainability measure $T_{pt}$ is regressed on $\mathit{PCI}_{pt}$ within each year cross-section, and the residuals $\hat{\varepsilon}_{pt}$ are retained. These error components are then standardised with a $Z$-transformation, resulting in two PCI-orthogonalised sustainability measures $\mathit{PSPI}_{pt}^\perp$ and $\mathit{PEPI}_{pt}^\perp$ which are uncorrelated with $\mathit{PCI}_{pt}$ by construction\footnote{Formally, the PCI-orthogonalised sustainability measures are constructed as $T_{pt}^\perp \equiv (\hat{\varepsilon}_{pt} - \mu(\hat{\varepsilon}_{pt})) / \sigma(\hat{\varepsilon}_{pt})$, where $\hat{\varepsilon}_{pt}$ is the fitted residual in the cross-sectional regression $T_{pt} = \alpha_{0t} + \alpha_{1t} C_{pt} + \varepsilon_{pt}$ for year $t$ with $\mathbb{E}_t[\varepsilon_{pt}] = 0$ and $\mathbb{E}_t[\varepsilon_{pt}C_{pt}] = 0$, ensuring orthogonality to product complexity.}. These orthogonalised measures are `cleansed' from linear covariation with product complexity, thereby isolating the effect of sustainability on diversification dynamics while suppressing standard error inflation. 

The full two-way interaction panel logit model is expressed as 
\begin{equation}\label{eq_logistic}
\begin{aligned}
    \mathbb{P}(\Delta M_{cpt} = 1) = 
     \Lambda \Big( & \delta_1 \omega_{cpt-1} + \delta_2 \log(\mathit{RCA}_{cpt-1}) \\
     & + \delta_3 \mathit{PCI}_{pt-1} + \delta_4 \mathit{PSPI}_{pt-1}^\perp + \delta_5 \mathit{PEPI}_{pt-1}^\perp \\
     & + \delta_6 (\omega_{cpt-1} \times \mathit{PCI}_{pt-1}) \\ 
     & + \delta_7 (\omega_{cpt-1} \times \mathit{PSPI}_{pt-1}^\perp) \\
     & + \delta_8 (\omega_{cpt-1} \times \mathit{PEPI}_{pt-1}^\perp) \\ 
     & + \lambda_{ct}
    \Big), \qquad \qquad \qquad \ \forall (c, p, t) : B_{cpt} = 1
\end{aligned} 
\end{equation}
% \begin{equation}\label{eq_logistic}
% {
% \begin{aligned}
%     \mathbb{P}(\Delta M_{cpt} = 1) = 
%      \Lambda \big( & \delta_1 \omega_{cpt-1} + \delta_2 \log(\mathit{RCA}_{cpt-1}) + \delta_3 \mathit{PCI}_{pt-1} + \delta_4 \mathit{PSPI}_{pt-1}^\perp \\
%      & + \delta_5 \mathit{PEPI}_{pt-1}^\perp + \delta_6 (\omega_{cpt-1} \times \mathit{PCI}_{pt-1}) \\
%      & + \delta_7 (\omega_{cpt-1} \times \mathit{PSPI}_{pt-1}^\perp) + \delta_8 (\omega_{cpt-1} \times \mathit{PEPI}_{pt-1}^\perp) \\
%      & + \lambda_{ct}
%     \big), \qquad \qquad \qquad \qquad \qquad \qquad \ \forall (c, p, t) : B_{cpt} = 1
% \end{aligned} 
% }
% \end{equation}
where $\Lambda(x) = 1 / (1 + \exp(x))$ is the logistic cumulative distribution function which converts the linear predictor into a probability bounded between $0$ and $1$. Next, $\omega_{cpt-1}$ denotes the relatedness density between country $c$ and product $p$ at time $t - 1$, which accounts for the capability structure present within $c$ in the previous period. The term $\log(\mathit{RCA}_{cpt-1})$ captures path dependence in comparative advantage, recognising that RCA values below unity might still signal a trajectory toward entry. $\mathit{PCI}_{pt-1}$ is the lagged product complexity index, $\mathit{PSPI}_{pt-1}^\perp$ is the lagged product social progress index, and $\mathit{PEPI}_{pt-1}^\perp$ is the lagged product environmental performance index, both orthogonalised with respect to $\mathit{PCI}_{pt-1}$. 

Furthermore, the two-way interactions enable the marginal effects of product complexity and sustainability to vary with relatedness, reflecting their nonlinear effect on diversification contingent on capability constraints. For instance, the effect of unrelated complex products ($\delta_3$) is expected to be negative, whereas the coefficient of related complex products ($\delta_6$) on diversification is anticipated to be positive. Contrasting these effects of complexity with those of the social and environmental sustainability indicators, the relative impact of complexity versus sustainability on the probability of diversification is evaluated. Given that $\mathit{PSPI}_{pt-1}^\perp$ and $\mathit{PEPI}_{pt-1}^\perp$ are orthogonalised with respect to $\mathit{PCI}_{pt-1}$, they capture the residual effects of social and environmental sustainability, disentangled from inherent knowledge intensity. This effectively tests whether sustainability has an \textit{additional} impact on the likelihood of entry, on top of the effect of complexity. For instance, if the estimated coefficient $\hat{\delta}_6$ is greater than $\hat{\delta}_7$ and/or $\hat{\delta}_8$, this would mean that countries are more likely to diversify into complex products per additional unit of relatedness, rather than in equally related socially or environmentally sustainable products (and \textit{vice versa})\todoh{this is not completely accurate: if delta_3 + delta_6 is greater, then this is the case? delta_6 represents the marginal impact of PCI per unit of relatedness}. If there is no tendency for countries to prioritise complex over sustainable products, these coefficients should be similar in sign and magnitude. Lastly, to account for potential omitted variable bias, country-year fixed effects $\lambda_{ct}$ are included, which capture time-varying unobserved heterogeneity across countries, such as changes in macroeconomic conditions, policy changes and shifts in complexity and sustainability outcomes. The model is estimated for all country-product-year combinations satisfying the backward condition ($B_{cpt} = 1$), ensuring that only observations are considered for which entry is theoretically possible (i.e. $\mathbb{P}(\Delta M_{cpt}) > 0$). First, the model is estimated on the full sample of \var{n_countries} countries, followed by separate estimations for high- and upper-middle-income countries versus lower-middle- and low-income countries to disentangle the role of income. 

\section{Results}\label{results}

Below, we present the results of the analysis. Section~\ref{c_lvl_results} examines the diversification opportunities available to countries, highlighting cross-country heterogeneity in the sustainability alignment of these prospects. Additionally, Section~\ref{p_lvl_results} exhibits the estimates of the logistic product entry model, which reveal that countries have a systematic tendency to diversify into related complex rather than related sustainable products. Notably, this effect is more pronounced for poorer than for richer countries. 

\subsection{Diversification Opportunities and Complexity-Sustainability Alignment}\label{c_lvl_results}

\begin{figure}[H]
    \centering
    \includegraphics[width=0.80\linewidth]{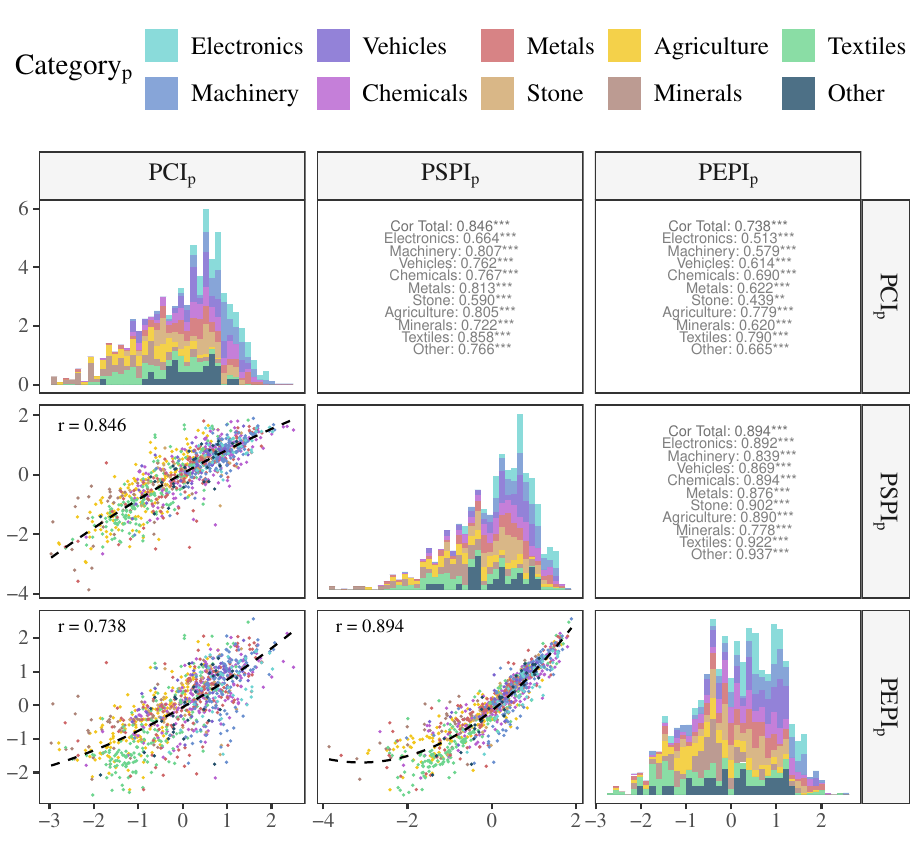}
    \caption{Distributions and Pearson Correlations of the Three Product Performance Indicators (\var{year_S_E_C_p_scat})}
    \label{fig_S_E_C_p_scat}
\end{figure}

Fig.~\ref{fig_S_E_C_p_scat} shows the distributions of and relationships among the three product performance scores: complexity ($\mathit{PCI}_p$), social ($\mathit{PSPI}_p$) and environmental ($\mathit{PEPI}_p$) performance in \var{year_S_E_C_p_scat}. The values in the upper-triangle show the global Pearson correlations by product category\footnote{The 19 HS sections are grouped into 10 broader categories for visualisation purposes only.}. 

The off-diagonal scatter plots show that product complexity is strongly and positively correlated with both social ($r = \var{cor_pci_p_spi_p_2022}$) and environmental performance ($r = \var{cor_pci_p_epi_p_2022}$), and that this pattern holds across all product categories. This indicates a strong and positive global alignment between product complexity and both sustainability indicators, meaning that more complex products are more likely to perform well in social and environmental indicators. Notably, compared with social performance, the $\mathit{PCI}_p$-$\mathit{PEPI}_p$ relationship is weaker and exhibits a slightly convex pattern. This nonlinear complexity-environmental connection is an important feature that drives the heterogeneity in \textit{local} opportunity alignment at the country-level. 

Next, the empirical distributions on the diagonal of Fig.~\ref{fig_S_E_C_p_scat} reveal that, on average, electronics ranks as the most complex category, followed by machinery, vehicles, chemicals, metals, stone, agriculture, minerals, and textiles as the least complex. As can be seen from the histograms, this ranking of the categories in terms of both social and environmental performance is largely preserved, suggesting that more complex products generally have higher $\mathit{PSPI}_p$ and $\mathit{PEPI}_p$ scores. However, the considerable overlap between the distributions of the various product categories indicates that there is greater variability of scores \textit{within} categories than \textit{between} categories (see variance decomposition in \ref{app_T_p_var_decomp}). % For instance, despite being the lowest ranking category, many textile products have greater $\mathit{PCI}_p$, $\mathit{PSPI}_p$ and $\mathit{PEPI}_p$ scores than a substantial number of electronic products. 

\begin{figure}[H]
    \centering
    \includegraphics[width=\linewidth]{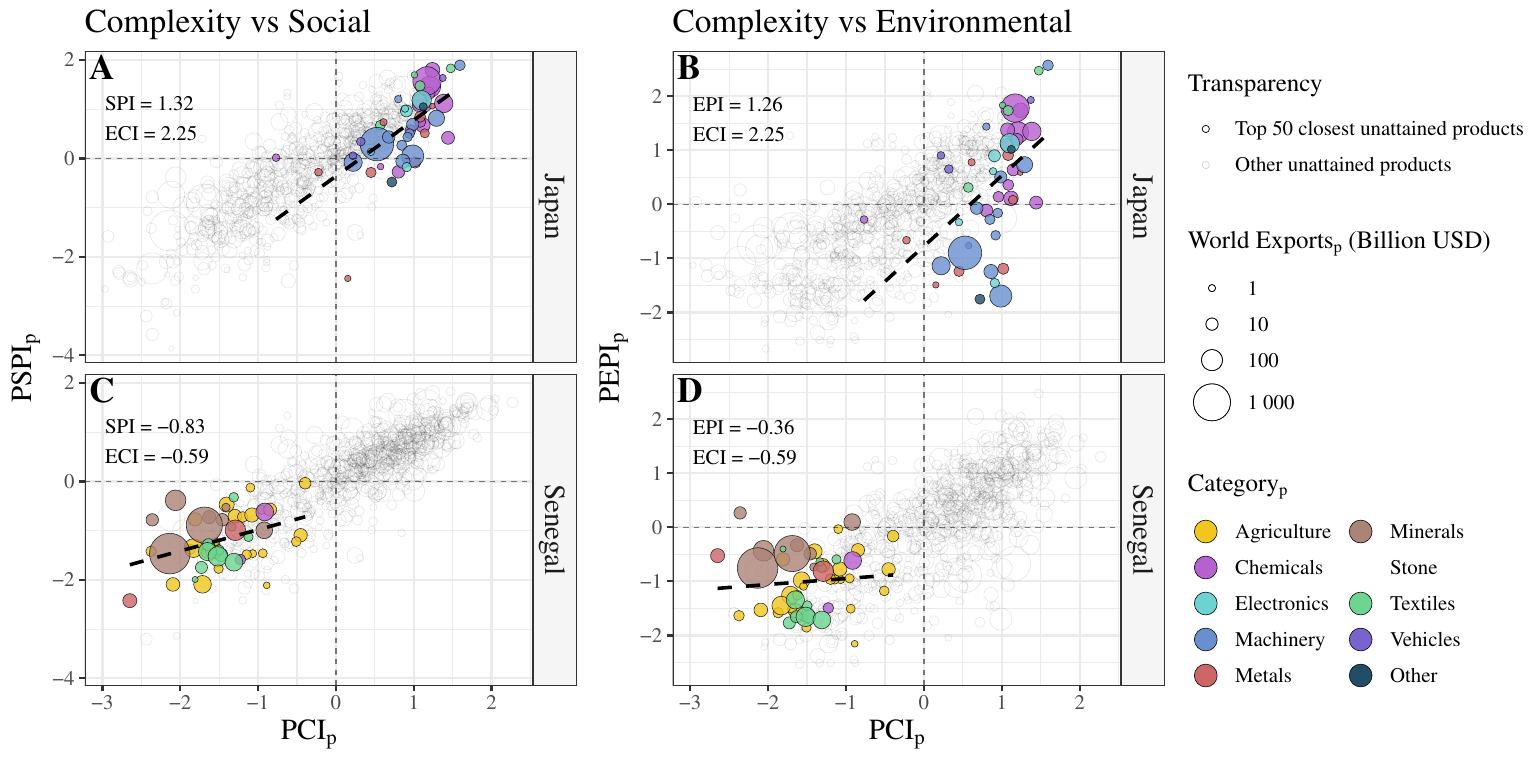}
    \caption{Top 50 Most Related Opportunities for \var{opp_plot_country_1} and \var{opp_plot_country_2} in Terms of Complexity vs Social Performance (A \& C) and Complexity vs Environmental Performance (B \& D) (\var{opp_plot_year})}
    \label{fig_opp_plot}
\end{figure}

To identify the most likely diversification opportunities for individual countries, these product performance measures are combined with information on relatedness density. Fig.~\ref{fig_opp_plot} shows the top $k = 50$ most related products for \var{opp_plot_country_1} and \var{opp_plot_country_2}, for which they do not yet have an $\mathit{RCA}_{cpt} \ge 1$ in \var{opp_plot_year} (i.e. $p \in \Omega_{ct}$), in terms of their complexity versus social performance on the left, and complexity versus environmental performance on the right. 

    % todoh: fix colour for missing categories: perhaps with get_legend from a plot where all categories are present
    
These opportunity plots show that \var{opp_plot_country_2} is closest to minerals, textiles and agricultural products, while \var{opp_plot_country_1} has more opportunities to diversify into machinery, chemicals and electronic products. Additionally, the joint distributions of these productive opportunities reveal two important differences between the two countries.

First of all, the products closest to \var{opp_plot_country_2} perform considerably worse across all three dimensions compared to the opportunities available to \var{opp_plot_country_1}. This suggests that \var{opp_plot_country_1} is positioned to transition into more complex, socially and environmentally sustainable products than \var{opp_plot_country_2}. Secondly, in addition to the differences in the average performance of these opportunities, a divergence is observed in the relationships \textit{between} the sustainability dimensions for \var{opp_plot_country_1} and \var{opp_plot_country_2}. For the opportunities shown in the social-complexity space on Fig.~\ref{fig_opp_plot}A \& \ref{fig_opp_plot}C, the slopes ($\hat{\beta}^{CS}_c$) of the point clouds are fairly similar for \var{opp_plot_country_1} and \var{opp_plot_country_2}. In contrast, the slopes of the products in the environmental-complexity space ($\hat{\beta}^{CE}_c$) on Fig.~\ref{fig_opp_plot}B \& \ref{fig_opp_plot}D are markedly different for both countries. For \var{opp_plot_country_1}, the correlation between the complexity and environmental performance of its closest products is high, allowing \var{opp_plot_country_1} to identify products that perform well in both dimensions simultaneously. Conversely, this slope is much weaker for \var{opp_plot_country_2}, suggesting that it will face more difficulties in finding products that are performant in both dimensions. In other words, this means that the arrow of expected development in the complexity-environmental plane points to the right for \var{opp_plot_country_2}, and diagonally upwards for \var{opp_plot_country_1}. 

\begin{figure}[H]
    \centering
    \includegraphics[width=0.8\linewidth]{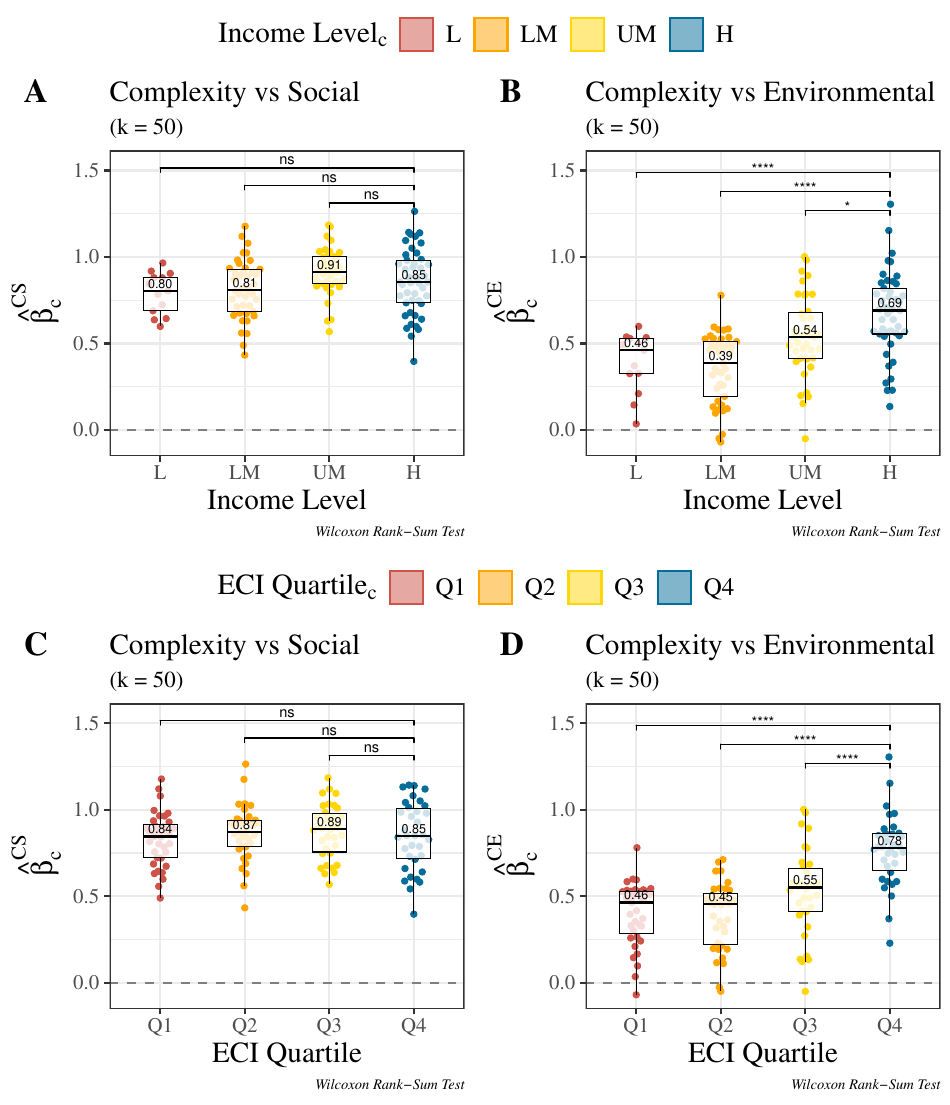}
    \caption{Distributions of the Opportunity Alignment Slopes by Income Level (A \& B) and by ECI quartile (C \& D) (\var{year_cor_dens})}
    \label{fig_slope_beeswarm}
\end{figure}

Fig.~\ref{fig_slope_beeswarm} shows the distributions of opportunity alignment scores across countries of the four income levels classified by \citet{TheWorldBank2023c} as high- (H), upper-middle- (UM), lower-middle- (LM), and low- (L) income countries (panels A \& B), and by ECI quartiles (panels C \& D). The values of these measures represent the expected gain in product sustainability associated with a unit increase in PCI for a product within the country's opportunity set $\Omega_{ct}$. Note that these slopes therefore exclusively capture information about the \textit{relationship between} PCI and the sustainability metrics, and not about the expected \textit{performance} on these measures. Consequently, the slopes reflect the \textit{direction} within the complexity-sustainability space in which countries are expected to move, given their most accessible opportunities. 

The figure reveals that, barring a small number of lower- and middle-income countries which faced environmentally misaligned opportunities in 2022 (Myanmar, Uzbekistan, Indonesia, and Laos), the two sustainability dimensions generally experience a positive connection with complexity during all stages of development. We find a highly similar pattern when considering ECI quartiles, which conforms with prior results in the literature. However, the slopes between complexity and social performance are, on average, steeper than for the environmental dimension. This reconfirms that complexity is more tightly aligned with social progress than with environmental performance. Additionally, we observe that while the complexity-social slopes do not vary significantly across income levels or ECI quartiles, the complexity-environmental slopes exhibit substantial heterogeneity. As indicated by the significant comparisons of the Wilcoxon rank-sum tests, more developed countries (high-income or high-ECI) face significantly more environmentally aligned diversification opportunities than less developed economies (middle- and low-income or lower-ECI countries). This result appears to be driven by the convex shape of the complexity-environmental relationship, indicating that the association between $\mathit{PCI}_p$ and $\mathit{PEPI}_p$ is weaker among lower-performing products but strengthens progressively for products at the higher end of the distribution. This nonlinear connection between product complexity and product environmental sustainability is reminiscent of the environmental Kuznets curve discussed in Section~\ref{complexity_and_sustainable_development}, suggesting that most diversification paths available to lower-income countries exhibit a considerably weaker complexity-environmental sustainability connection. By comparison, the lack of significant variation in complexity-social slopes across development levels suggests a consistent alignment between product complexity and social performance.

\subsection{Diversification Bias Towards Complexity}\label{p_lvl_results}

Table~\ref{modB} presents the estimation results of equation~(\ref{eq_logistic}), which models the probability of entry as a function of lagged product characteristics and their interactions with relatedness density. As discussed, the $\mathit{PSPI}_p$ and $\mathit{PEPI}_p$ are orthogonalised with respect to $\mathit{PCI}_p$ to avoid multicollinearity issues (see comparison of the variance inflation factors in \ref{app_vifs})\footnote{As a robustness check, we also report the results when $\mathit{PSPI}$ and $\mathit{PEPI}$ are not orthogonalised with respect to $\mathit{PCI}$. The sign and significance of all coefficients remain qualitatively unchanged, except for $\hat{\delta}_8$ which loses significance for H \& UM income countries (see \ref{app_modB_notorth}).}. Column (1) shows the estimates for the full sample without interaction terms, while column (2) includes them. Thereafter, columns (3) and (4) report the estimates for high- and upper-middle- (H \& UM) income countries versus lower-middle- and low- (LM \& L) income countries, respectively. All estimations are restricted to rows for which the backward condition is fulfilled, and include country-time fixed effects to control for any country-specific shocks, policy changes, or time-varying macroeconomic trends. 

\begin{table}[H]
   \caption{\label{modB} Product Entry Model Estimates}
   \vspace{6pt}
   \footnotesize
   
   \begin{tabular}{lcccc}
      \toprule
      Dependent Variable: & \multicolumn{4}{c}{$\mathbb{{P}}(\Delta M_{cpt} = 1)$}\\
       & \multicolumn{2}{c}{Full Sample} & H \& UM & LM \& L \\ 
      Model:                                                       & (1)           & (2)            & (3)            & (4)\\  
      \midrule
      \emph{Variables}\\
      $\omega_{cpt-1}$                                             & 8.398$^{***}$ & 7.706$^{***}$  & 6.269$^{***}$  & 10.105$^{***}$\\   
                                                                   & (0.795)       & (0.782)        & (0.872)        & (2.111)\\   
      $\log(\mathit{RCA}_{cpt-1})$                                 & 1.646$^{***}$ & 1.638$^{***}$  & 2.014$^{***}$  & 1.206$^{***}$\\   
                                                                   & (0.089)       & (0.089)        & (0.149)        & (0.085)\\   
      $\mathit{PCI}_{pt-1}$                                        & -0.047        & -0.298$^{***}$ & -0.240$^{***}$ & -0.457$^{***}$\\   
                                                                   & (0.029)       & (0.060)        & (0.066)        & (0.104)\\   
      $\mathit{PSPI}^{\perp}_{pt-1}$                               & -0.027        & 0.011          & 0.017          & -0.061\\   
                                                                   & (0.032)       & (0.053)        & (0.066)        & (0.091)\\   
      $\mathit{PEPI}^{\perp}_{pt-1}$                               & 0.021         & -0.072         & -0.080         & 0.026\\   
                                                                   & (0.035)       & (0.058)        & (0.072)        & (0.097)\\   
      $\omega_{cpt-1}$ $\times$ $\mathit{PCI}_{pt-1}$              &               & 1.062$^{***}$  & 0.841$^{***}$  & 1.917$^{***}$\\   
                                                                   &               & (0.206)        & (0.224)        & (0.530)\\   
      $\omega_{cpt-1}$ $\times$ $\mathit{PSPI}^{\perp}_{pt-1}$     &               & -0.167         & -0.249         & 0.476\\   
                                                                   &               & (0.193)        & (0.231)        & (0.452)\\   
      $\omega_{cpt-1}$ $\times$ $\mathit{PEPI}^{\perp}_{pt-1}$     &               & 0.429$^{**}$   & 0.535$^{**}$   & -0.429\\   
                                                                   &               & (0.215)        & (0.263)        & (0.454)\\   
      \midrule
      \emph{Fixed-effects}\\
      $c$-$t$                                                      & Yes           & Yes            & Yes            & Yes\\  
      \midrule
      \emph{Fit statistics}\\
      Observations                                                 & 728,978       & 728,978        & 498,034        & 229,881\\  
      Countries                                                    & 137           & 137            & 89             & 61\\  
      Years                                                        & 10            & 10             & 10             & 10\\  
      Pseudo R$^2$                                                 & 0.244         & 0.245          & 0.245          & 0.257\\  
      ROC-AUC                                                      & 0.906         & 0.906          & 0.909          & 0.909\\  
      F$_1$                                                        & 0.178         & 0.177          & 0.182          & 0.189\\  
      \bottomrule
      \multicolumn{5}{l}{\emph{Clustered ($c$ \& $p$) standard-errors in parentheses}}\\
      \multicolumn{5}{l}{\emph{Signif. Codes: ***: 0.01, **: 0.05, *: 0.1}}\\
   \end{tabular}
\end{table}

First, the goodness of fit statistics of these models should be judged in light of the fact that sustained diversification as defined in this analysis ($\Delta M_{cpt} = 1$) is a highly rare event. Among the \var{eligible_rows_count} observations eligible for diversification (i.e. for which the backward condition is met), only \var{new_M_cp_count} cases (\var{new_M_cp_prevalence}~\%) of sustained entry are observed. Nevertheless, the McFadden pseudo $R^2$ values above $0.24$ are generally regarded to signal a strong goodness of fit in the context of logistic regression, given the skewed class distribution \citep{McFadden1977}. Additionally, although the severe class imbalance weakens threshold-dependent predictive accuracy measures such as the $F_1$ score, inference remains valid due to a reliable ranking of the products as indicated by the high areas under the receiver operator characteristic curves (ROC-AUCs, see \ref{app_combined_rocs}) \citep{Hanley.McNeil1983}. 

As expected, relatedness density is by far the most important predictor of product appearances across all specifications. However, it should be noted that the large coefficients of relatedness correspond to an increase in the log-odds of entry following a \textit{unit-increase} in relatedness density. Considering that $\omega_{cpt} \in [0, 1]$, this is an extrapolation that never occurs in reality. More practically, the full-sample estimate in column (1) implies that a $0.10$ point increase in $\omega_{cpt-1}$ raises the log-odds of entry in year $t$ by $\var{mC_1_tpi_ALL_dens_logodds} \times 0.10$, which corresponds to an odds ratio of $\exp(\var{mC_1_tpi_ALL_dens_logodds_0.1}) \approx \var{mC_1_tpi_ALL_dens_oddsratio}$. In other words, a $10$ percentage-point increase in relatedness roughly doubles the odds of entry, \textit{ceteris paribus}. Similarly, all specifications show a robust, positive effect of log RCA on the probability of entry, underlining the path-dependent nature of export expansion. Furthermore, the baseline results show that, on their own, the three product performance measures do not significantly impact the likelihood of diversification, all other things held constant. 

Specification (2) accounts for the nonlinear effect that these product characteristics might have on the probability of diversification, conditional upon the capabilities present within countries. By introducing these interaction terms, it now becomes clear that the effects of product complexity and sustainability are different for related versus unrelated products. Now, the direct effect of $\mathit{PCI}_{pt-1}$ is significantly negative, which means that unrelated complex products are harder to enter, all else being equal. In contrast, when interacted with relatedness, this effect becomes positive, which means that having sufficient capabilities to reach a more complex product significantly increases product entries. The relatedness cut-off at which the marginal effect of complexity on the log-odds of entry becomes positive is given by
\begin{equation}
    \frac{\partial \ln\left(\frac{\mathbb{P}}{1 - \mathbb{P}}\right)}{\partial \mathit{PCI_{cpt-1}}} = \delta_3 + \delta_6 \omega_{cpt-1} > 0
\end{equation} 
where $\mathbb{P} = \mathbb{P}(\Delta M_{cpt} = 1)$, and which corresponds to a threshold of $\omega^{\star}_\mathit{PCI} = -\delta_3 / \delta_6$. Specifically, the estimates suggest that a higher $\mathit{PCI}_{cpt-1}$ only raises the odds of entry once relatedness density is greater than $-(\var{mC_2_tpi_ALL_delta_3} / \var{mC_2_tpi_ALL_delta_6}) \approx \var{mC_2_tpi_ALL_dens_threshold_pci}$. This is a relatively high barrier, considering that there were only $\var{n_countries_above_mC_2_tpi_ALL_dens_threshold_pci}$ countries in \var{logistic_threshold_year} that had at least $1$ unattained product with a density above this threshold. 

In addition, while there is no significant direct effect of environmental performance when stripped of complexity, the coefficient of the interaction term with relatedness is significantly larger than zero. The relatedness threshold at which a $1$ standard deviation increase in $\mathit{PEPI}_{pt-1}^\perp$ raises the odds of entry is $-(\var{mC_2_tpi_ALL_delta_5} / \var{mC_2_tpi_ALL_delta_8}) \approx \var{mC_2_tpi_ALL_dens_threshold_epi}$. Notably, this threshold is considerably lower than the relatedness threshold for $\mathit{PCI}_{cpt-1}$, which means that countries are more likely to diversify into more environmentally sustainable products at much lower capability requirements. Besides this, $\mathit{PSPI}_{pt-1}^\perp$ has neither a direct nor interaction effect that reaches significance, suggesting that variation in a product's social progress does not drive countries' entry decisions once product complexity is taken into account. 

Finally, columns (3) and (4) show the results for H \& UM versus LM \& L income countries, respectively. This comparison yields two interesting results. First, lower-income economies exhibit both a lower stand-alone effect of $\mathit{PCI}_{cpt-1}$, and a complexity-relatedness interaction effect that is more than double than that observed in richer countries. This means that poorer countries face comparatively steeper capability barriers to diversify into complex products, whereas wealthier nations can enter complex industries at lower levels of relatedness. On the other hand, the amplifying effect of complexity on entry per additional unit of relatedness is much larger for these less developed economies. Nevertheless, the critical relatedness density after which $\mathit{PCI}_{cpt-1}$ boosts entry is relatively similar across both groups ($\var{mC_2_tpi_HUM_dens_threshold_pci}$ for richer and $\var{mC_2_tpi_LML_dens_threshold_pci}$ for poorer countries), which reinforces the notion that capability constraints drive the development of complex industries in a comparable way across income levels. Considering that in \var{logistic_threshold_year}, the average relatedness density of the top $50$ closest unattained products ($p \in \Omega_{ct}$) was $\var{avg_dens_cp_topk50_HUM}$ for H \& UM income countries but only $\var{avg_dens_cp_topk50_LML}$ for LM \& L income countries, this sheds light on why diversification into complex products remains difficult for developing nations. 

% This means that LML have difficulties moving to the right as well

Second, a strong disparity is observed in the effect of the environmental-relatedness interaction term across both income groups. The coefficient of this interaction effect represents the mixed second-order partial derivative 
\begin{equation} \label{eq_second_order_partial_derivative}
    \delta_8 = \frac{\partial^2 \ln\left(\frac{\mathbb{P}}{1 - \mathbb{P}}\right)}{\partial \mathit{PEPI}_{pt-1} \partial \omega_{cpt-1}}
\end{equation}
which is positive and statistically significant for H \& UM income countries ($\hat{\delta}_8^\mathrm{HUM} = \var{mC_2_tpi_HUM_delta_8}$), but negative and insignificant for LM \& L income countries ($\hat{\delta}_8^{\mathrm{LML}} = \var{mC_2_tpi_LML_delta_8}$). This implies that for richer countries, being close to an environmentally sustainable product \textit{increases} the odds of entry, while for developing economies, related sustainable products are not more likely to be diversified into, \textit{ceteris paribus}. Put differently, while capability upgrades in richer economies promote green diversification alongside complexity gains, this is not true for lower-income countries. Instead, their gains in relatedness are expected to stimulate diversification in complex products, rather than environmentally sustainable ones. Crucially, the insignificant effect of $\mathit{PEPI}_{pt-1}$ for lower-income countries does not rule out the possibility of entry into simultaneously complex and green products. After all, as shown in Fig.~\ref{fig_slope_beeswarm}, product complexity and environmental performance are still positively aligned for these countries, meaning that movements towards complex industries might still coincide with environmental gains. Once again, no significant effects were detected for $\mathit{PSPI}^\perp_{pt-1}$ for either income group. 

\begin{figure}[H]
    \centering
    \includegraphics[width=\linewidth]{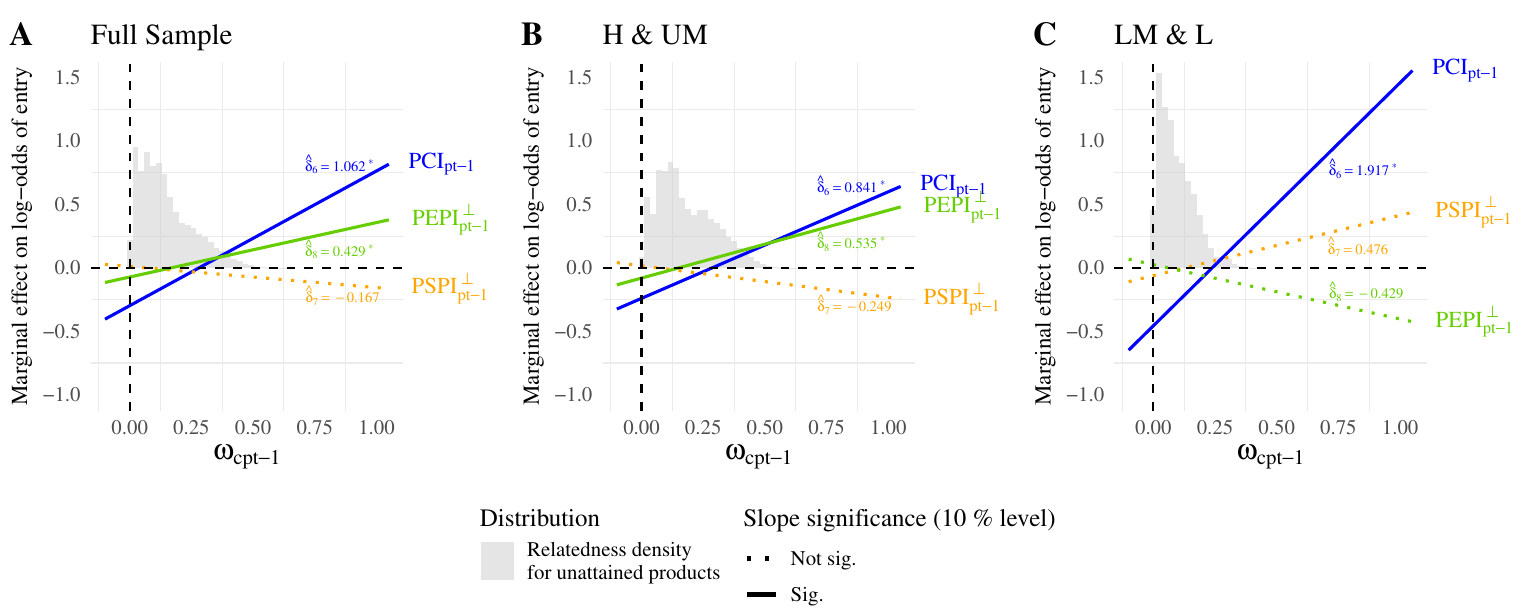}
    \caption{Marginal Effects of $\mathit{PCI}_{pt-1}$, $\mathit{PSPI}_{pt-1}^{\perp}$, and $\mathit{PEPI}_{pt-1}^\perp$ on the Log-Odds of Entry, Conditional on Relatedness Density}
    % \caption*{\textit{Note:} Empirical distribution of relatedness density }
    \label{fig_logistic_persistent_marginal_effects_line}
\end{figure}

Fig.~\ref{fig_logistic_persistent_marginal_effects_line} visually displays the marginal effects of $\mathit{PCI}_{pt-1}$, $\mathit{PSPI}_{pt-1}^\perp$ and $\mathit{PEPI}_{pt-1}^\perp$ on the log-odds of entry, conditional on the value of relatedness density for both the full (Fig.~\ref{fig_logistic_persistent_marginal_effects_line}A) and subsamples (Fig.~\ref{fig_logistic_persistent_marginal_effects_line}B and \ref{fig_logistic_persistent_marginal_effects_line}C). The slopes of these lines correspond to the second-order partial derivatives represented by the coefficients $\hat{\delta}_6$, $\hat{\delta}_7$ and $\hat{\delta}_8$ as formulated in equation~(\ref{eq_second_order_partial_derivative}), and the vertical intercepts represent the direct effects of each variable captured by $\hat{\delta}_3$, $\hat{\delta}_4$ and $\hat{\delta}_5$. Solid lines represent significant interaction effects, while dotted lines denote insignificant effects at the $10~\%$ confidence level. The grey histograms show the empirical distributions of relatedness density for unattained products ($M_{cpt} = 0$) for each sample. 

\begin{figure}[H]
    \centering
    \includegraphics[width=\linewidth]{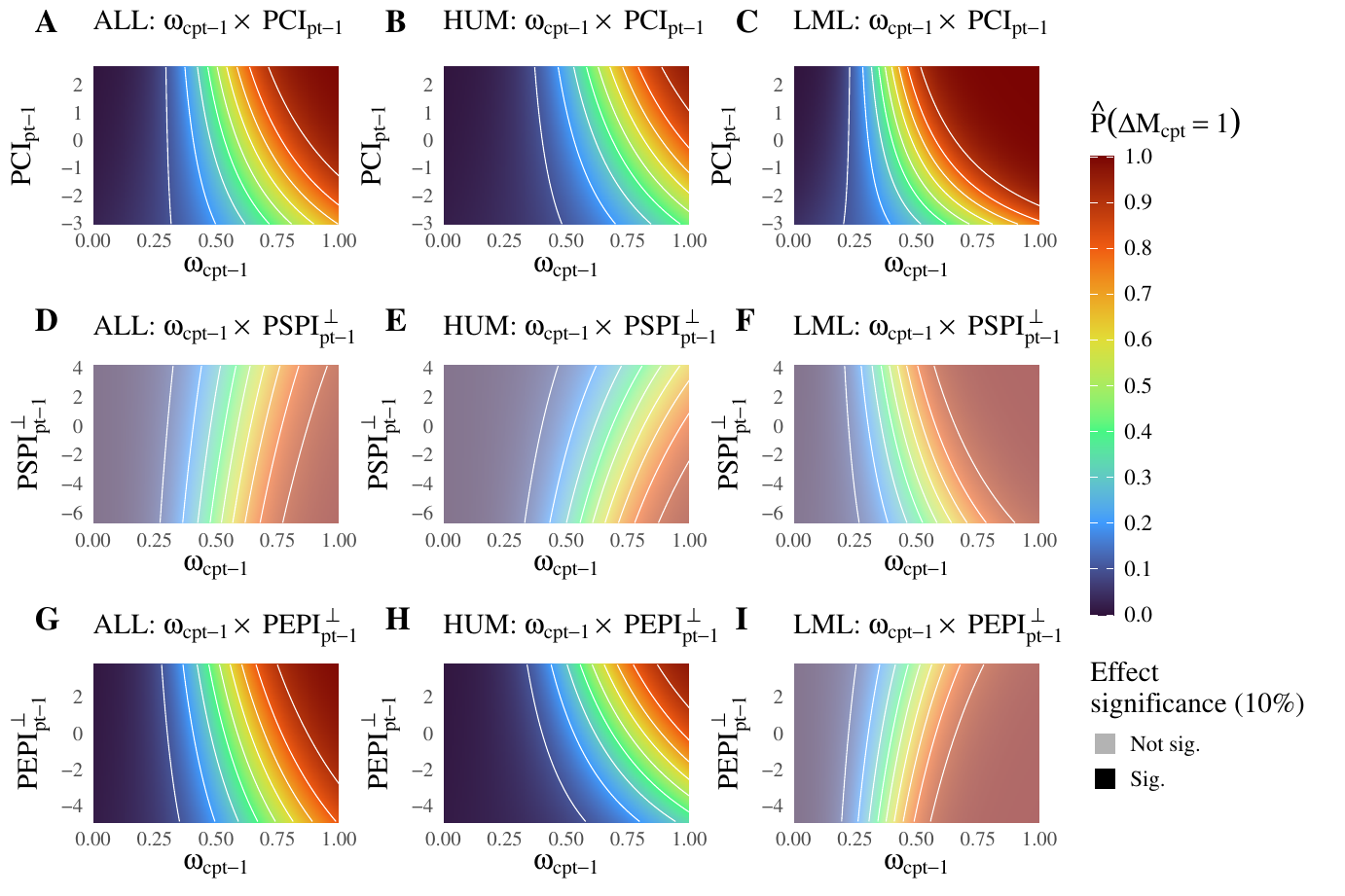}
    \caption{Joint Effect of Relatedness and Product Performance on Entry}
    \caption*{\textit{Note:} Colour represents the predicted probability of entry conditional on relatedness density $\omega_{cpt-1}$ and product performance ($\mathit{PCI}_{pt-1}$, $\mathit{PSPI}^\perp_{pt-1}$ or $\mathit{PEPI}^\perp_{pt-1}$), with $\log(\mathit{RCA}_{cpt-1})$ held at its median, all other variables held at their mean, and fixed effects excluded. Note that the effect of $\mathit{PSPI}^\perp_{pt-1}$ is insignificantly different from $0$ at the $10~\%$ confidence level.}
    \label{fig_pred_new_M_cp_heatmap}
\end{figure}

These figures illustrate that the effect of $\mathit{PCI}_{pt-1}$ is negative until density reaches a certain threshold, after which complexity increases the odds of entry. Next to that, the figure clearly shows how the marginal effect of $\mathit{PEPI}_{pt-1}^\perp$ rises with relatedness for H \& UM income countries (Fig.~\ref{fig_logistic_persistent_marginal_effects_line}B), but does not significantly impact entry for LM \& L income countries (Fig.~\ref{fig_logistic_persistent_marginal_effects_line}C). Fig.~\ref{fig_pred_new_M_cp_heatmap} confirms these results by using the estimated coefficients of the model to plot the predicted probability of entry $\hat{\mathbb{P}}(\Delta M_{cpt} = 1)$ as a function of lagged relatedness density and lagged product performance measures ($\mathit{PCI}_{pt-1}$, $\mathit{PSPI}^\perp_{pt-1}$ and $\mathit{PEPI}^\perp_{pt-1}$), while $\log(\mathit{RCA}_{cpt-1})$ is kept at its median, all other variables are held at their mean, and fixed effects are excluded. Panels A, D and G show the results for the full sample (ALL); B, E and H for high- and upper-middle-income countries (HUM); and C, F and I for lower-middle- and lower-income countries (LML). Transparent panels indicate effects are calculated using statistically insignificant coefficients at the $10~\%$ confidence level. 

 % todo: reduce font size of notes? 

Again, this figure clearly shows the nonlinear effect of $\mathit{PCI}_{pt-1}$ on the probability of entry, depending on the level of relatedness density (panels A--C). While at low levels of relatedness, product complexity decreases the odds of entry, this effect becomes positive for more related opportunities. As argued before, this effect seems more pronounced for less developed economies. 

Panels G--I reaffirm the asymmetric effect of product environmental performance on the odds of entry for countries of different income groups. While for richer countries, the odds of entry significantly increase with $\mathit{PEPI}^\perp_{pt-1}$, this effect is reversed and insignificant for poorer economies. A similar but opposite asymmetry can be seen for $\mathit{PSPI}^\perp_{pt-1}$ in panels D--F. However, it should be noted that the estimated effect of $\mathit{PSPI}^\perp_{pt-1}$ is not significantly different from $0$ at the $10~\%$ confidence level. 

Lastly, the empirical distributions of relatedness density of the respective subsamples should be taken into account when interpreting these figures. For instance, while for H \& UM income countries densities range beyond $0.50$, these range only up to about $0.30$ for LM \& L income countries (see grey histograms in Fig.~\ref{fig_logistic_persistent_marginal_effects_line}). As a result, the predicted probabilities beyond these relatedness values are extrapolations that are not realised under feasible conditions. 

\section{Discussion}\label{discussion}

% todo: mention that these LML countries are the ones with more misaligned opportunities 

This study contributes to the expanding literature on the link between trade diversification and sustainable development by exploring product-level diversification dynamics. By modelling the nonlinear effect of product complexity, social, and environmental sustainability on the likelihood of product entry, the results offer three central insights with theoretical and policy relevance.  

First, we exploit country-level variation in the Social Progress Index and Environmental Performance Index to construct two measures for product-level social and environmental sustainability: the Product Social Progress Index and the Product Environmental Performance Index, respectively. Consistent with the existing literature, we find that product complexity is strongly positively associated with both social and environmental product performance, which corroborates positive global alignment between complexity and sustainability. 

Second, we investigate the specific diversification opportunities available to countries by combining these product sustainability indicators with product complexity and relatedness density. From this analysis, we create two opportunity alignment measures that capture the correspondence between the complexity and the social ($\hat{\beta}^{CS}_c$) versus environmental ($\hat{\beta}^{CE}_c$) sustainability of the closest diversification opportunities available to countries. Substantial variation is observed in the complexity-environmental slopes across income levels, with high-income countries facing significantly more diversification opportunities that are simultaneously complex and environmentally performant. Meanwhile, the connection between product complexity and product social progress is generally stronger and more consistent across income groups. 

Third, the granular analysis of product-level diversification dynamics provides evidence that countries are more likely to enter related complex products over related sustainable alternatives. On average, we find that for highly related products, complexity has a more dominant impact on the odds of entry than both social and environmental sustainability. This strong complexity-relatedness interaction effect reaffirms the intuition that the accumulation of productive capabilities drives the development of more complex industries. On top of that, it is observed that environmental sustainability has a small but positive contribution to product entry, once a basic level of relatedness is achieved. Interestingly, this effect is significantly positive for H \& UM income countries, but insignificant for LM \& L income countries. Put differently, richer economies have a systematic tendency to develop industries that are both complex and environmentally performant, while developing countries prioritise related complex industries over related environmentally sustainable ones. This might be due to the fact that higher-income countries possess the institutional capacity and resources needed to pursue more complex development goals, especially in the area of sustainability. No significant effects are found for social sustainability, suggesting that product social progress has no \textit{additional} impact on entry beyond the influence of complexity, regardless of the country's income level. 

That being said, this does not necessarily imply that developing countries deliberately prefer complexity at the expense of sustainability, but rather that there is a systematic tendency to venture into complex industries, conditional upon possessing the relevant capabilities. Additionally, given the strong positive correlations between PCI, PSPI and PEPI, entry in complex products will often coincide with sustainable development. These results rather point to the fact that at an identical level of product complexity and relatedness, environmentally sustainable products are more likely to be entered by high- than low-income countries. This again confirms the second insight of the paper that the complexity-environmental alignment of the diversification opportunities of developing economies is consistently lower than for richer nations. 

% Indeed, the model does not predict what products countries will move into, but rather explores how the product characteristics impact the likelihood of entry. 

% Because of the positive global relationship between PCI, PSPI and PEPI, diversification into complex p is expected to coincide with entry into sustainable p. However, when partialling out this positive covariance, we observe that complexity plays a more dominant role on the trajectories of countries than social and environmental product performance. Nevertheless, environmental sustainability has an additional impact on diversification beyond complexity, which ... 

It is important to note that the alignment measures do not capture the average performance of the opportunities available to a country, but merely the \textit{relationship between} their complexity and sustainability. As such, these slopes reflect the expected direction within the complexity-sustainability plane along which countries are most likely to diversify. Hence, while these patterns point to an inclination of development towards complexity over sustainability, a detailed examination of product sustainability impacts may reveal viable pathways toward export baskets that align all three objectives. 

Furthermore, the magnitudes of the effects of the product characteristics are small in comparison to the dominant role played by relatedness density and RCA, which are the principal determinants of future diversification trajectories. This path-dependence, alongside the suboptimal opportunity sets available to less developed economies, strengthens the argument supporting diversification into unrelated products. Such unrelated diversification or leapfrogging strategies are often argued to be essential for structural transformation, by accelerating economic catch-up \citep{Alshamsi.etal2018, Cimoli.etal2009, Hidalgo.etal2007, Lee2013}, enabling shifts in local capabilities \citep{Neffke.etal2018} and fostering radical innovation \citep{Castaldi.etal2015}. This would entail that countries should invest in developing the productive capabilities required to enter industries that are both highly complex and more sustainable. Nonetheless, as implied by the principle of relatedness, leapfrogging is an uncommon and demanding process, as it typically necessitates the acquisition of a broad array of new capabilities \citep{Hausmann.Hidalgo2011, Hidalgo.etal2007, Pinheiro2024}. This is further complicated by the fact that the relatedness algorithm's agnosticism regarding the precise nature of these capabilities makes it challenging to identify the specific requirements for developing new industries \citep{Hidalgo2021}. 

Finally, the interpretation of these results is contingent on the definition of the product sustainability indicators $\mathit{PSPI}_p$ and $\mathit{PEPI}_p$. The primary advantage of this weighted average-based method is that it allows us to compute highly granular, continuous product performance scores from country-level sustainability indicators and the specialisation network, without a pre-defined classification of ``green'' or ``social'' products. Despite this, this estimation method does not control for country characteristics and does not take advantage of the full panel data. For these reasons, an important avenue for further research is to develop an improved method to estimate the impact of individual product exports on country-level sustainability which accommodates these limitations. A second interesting avenue for further research could focus on individual social and environmental subdimensions. Decomposing the SPI and EPI into their respective components would allow us to evaluate the connection between product exports and specific subindicators of social and environmental sustainability, and explore interactions between them.

\section{Conclusion}

This paper contributes to a better understanding of the link between trade diversification and sustainable development. Building on the idea of complexity and relatedness, we explore the alignment between diversification geared towards related and complex products with two product-level measures for social and environmental sustainability. This analysis yields three main findings. 

First, consistent with the existing literature, we find that product complexity is generally strongly correlated with both social and environmental product performance.

Second, we observe that the diversification opportunities available to countries differ significantly depending on their income level. While the alignment between product complexity and product social progress is fairly consistent across income groups, the alignment between product complexity and product environmental progress is significantly higher for high-income countries. Higher-income countries, therefore, face more diversification opportunities that are both complex and environmentally performant, compared to lower-income countries. For the complexity-social alignment, such income differences are not observed.

Third, the results show that countries are systematically more likely to enter related complex over social or environmentally sustainable products, reaffirming the intuition that the accumulation of productive capabilities drives the development of more complex industries. On top of that, product environmental performance has a smaller but positive effect on product entry for high- and upper-middle-income countries once a basic level of relatedness is achieved, but no effect for lower-middle- and low-income countries. Richer countries are, in other words, more likely to diversify into proximate products that are simultaneously complex and environmentally performant, while developing countries tend to prioritise related complex products over related sustainable ones.

Importantly, this suggests that simultaneous progress across all three dimensions of sustainable development is possible, and to some extent arises spontaneously, due to the strong global alignment between complexity, environmental and social outcomes. However, lower-income countries do not experience the same levels of environmental alignment within their closest opportunities as more developed economies, which may limit their ability to achieve fast sustainable development. This presents an interesting puzzle for developing economies. On the one hand, one could argue that developing countries should pursue diversification strategies that are explicitly social and environmentally friendly. On the other hand, one might maintain that it is reasonable for developing economies to focus on raising their complexity and income, since this will move them towards a state where these goals are better aligned with sustainability objectives. This second option seems reasonable in a world where developing countries have limited institutional capacities. Efforts to micromanage economies to achieve multiple goals while sacrificing growth may backfire, achieving none of the intended objectives and leaving economies in an even more fragile condition. By building complexity, economies are likely to increase their income while growing the diversity of capacities needed to pursue more challenging development goals. 

%%%% Appendices %%%% 
\clearpage

% \section*{Data Availability}

% The authors do not have permission to share data.

\appendix

\section{Variance Decomposition of Product Performance Measures Across Categories}\label{app_T_p_var_decomp}

% Reset counters (to get A.1 instead of A.3)
\setcounter{table}{0} 
\setcounter{figure}{0} 
\setcounter{equation}{0} 

Table~\ref{tab:T_p_var_decomp} shows the decomposition of within versus between category variability of the three product performance measures $T_p$ in \var{opp_plot_year} using a one-way analysis of variance (ANOVA). SS represents the sum of squared errors, and MS is the mean squared error. $\eta^2 = \mathit{SS}_\mathrm{between} / \mathit{SS}_\mathrm{total}$ represents the share of total variance explained by category, and $\omega^2 = \frac{\mathit{SS}_\mathrm{between} - \mathit{df}_\mathrm{between} \cdot \mathit{MS}_\mathrm{error}}{\mathit{SS}_\mathrm{total} + \mathit{MS}_\mathrm{error}}$ is a bias-corrected estimate of the population share of variance explained by category. 

\begin{table}[H]

\caption{\label{tab:T_p_var_decomp}Variance Decomposition of Product Performance Variables by Product Category (2022)}
\footnotesize
\begin{tabular}[t]{lrrrrrr}
\toprule
Variable & $\mathit{SS}_\mathrm{between}$ & $\mathit{SS}_\mathrm{within}$ & \% between & \% within & $\eta^2$ & $\omega^2$\\
\midrule
$\mathit{PCI}_{p}$ & 445.166 & 582.834 & 0.433 & 0.567 & 0.433 & 0.428\\
$\mathit{PSPI}_{p}$ & 267.354 & 760.646 & 0.260 & 0.740 & 0.260 & 0.253\\
$\mathit{PEPI}_{p}$ & 211.296 & 816.704 & 0.206 & 0.794 & 0.206 & 0.198\\
\bottomrule
\end{tabular}
\end{table} 

For all three variables, the majority of the variance is explained by heterogeneity within rather than between categories. However, this proportion of within-category variability is considerably greater for $\mathit{PSPI}_p$ and $\mathit{PEPI}_p$ than for $\mathit{PCI}_p$.

\newpage
\section{Results of Product Entry Model without Orthogonalisation}\label{app_modB_notorth}

% Reset counters (to get A.1 instead of A.3)
\setcounter{table}{0} 
\setcounter{figure}{0} 
\setcounter{equation}{0} 

Table~\ref{modB_notorth} displays the estimates of the product entry model \textit{without} orthogonalising $\mathit{PSPI}_{pt-1}$ and $\mathit{PEPI}_{pt-1}$ with respect to $\mathit{PCI}_{pt-1}$. 

\begin{table}[H]
   \caption{\label{modB_notorth} Product Entry Model Estimates without Orthogonalisation}
   \vspace{6pt}
   \footnotesize
   
   \begin{tabular}{lcccc}
      \toprule
      Dependent Variable: & \multicolumn{4}{c}{$\mathbb{{P}}(\Delta M_{cpt} = 1)$}\\
       & \multicolumn{2}{c}{Full Sample} & H \& UM & LM \& L \\ 
      Model:                                              & (1)           & (2)           & (3)           & (4)\\  
      \midrule
      \emph{Variables}\\
      $\omega_{cpt-1}$                                    & 8.406$^{***}$ & 7.719$^{***}$ & 6.279$^{***}$ & 10.106$^{***}$\\   
                                                          & (0.795)       & (0.781)       & (0.870)       & (2.108)\\   
      $\log(\mathit{RCA}_{cpt-1})$                        & 1.646$^{***}$ & 1.638$^{***}$ & 2.014$^{***}$ & 1.206$^{***}$\\   
                                                          & (0.089)       & (0.089)       & (0.149)       & (0.085)\\   
      $\mathit{PCI}_{pt-1}$                               & -0.030        & -0.222$^{**}$ & -0.163$^{*}$  & -0.394$^{**}$\\   
                                                          & (0.049)       & (0.087)       & (0.099)       & (0.170)\\   
      $\mathit{PSPI}_{pt-1}$                              & -0.048        & 0.020         & 0.025         & -0.111\\   
                                                          & (0.059)       & (0.101)       & (0.124)       & (0.172)\\   
      $\mathit{PEPI}_{pt-1}$                              & 0.030         & -0.118        & -0.125        & 0.037\\   
                                                          & (0.056)       & (0.094)       & (0.118)       & (0.157)\\   
      $\omega_{cpt-1}$ $\times$ $\mathit{PCI}_{pt-1}$     &               & 0.773$^{**}$  & 0.546         & 1.741$^{**}$\\   
                                                          &               & (0.343)       & (0.376)       & (0.813)\\   
      $\omega_{cpt-1}$ $\times$ $\mathit{PSPI}_{pt-1}$    &               & -0.301        & -0.444        & 0.902\\   
                                                          &               & (0.365)       & (0.436)       & (0.850)\\   
      $\omega_{cpt-1}$ $\times$ $\mathit{PEPI}_{pt-1}$    &               & 0.685$^{*}$   & 0.842$^{*}$   & -0.725\\   
                                                          &               & (0.352)       & (0.430)       & (0.732)\\   
      \midrule
      \emph{Fixed-effects}\\
      $c$-$t$                                             & Yes           & Yes           & Yes           & Yes\\  
      \midrule
      \emph{Fit statistics}\\
      Observations                                        & 728,978       & 728,978       & 498,034       & 229,881\\  
      Countries                                           & 137           & 137           & 89            & 61\\  
      Years                                               & 10            & 10            & 10            & 10\\  
      Pseudo R$^2$                                        & 0.244         & 0.245         & 0.245         & 0.257\\  
      ROC-AUC                                             & 0.906         & 0.906         & 0.909         & 0.909\\  
      F$_1$                                               & 0.178         & 0.177         & 0.182         & 0.189\\  
      \bottomrule
      \multicolumn{5}{l}{\emph{Clustered ($c$ \& $p$) standard-errors in parentheses}}\\
      \multicolumn{5}{l}{\emph{Signif. Codes: ***: 0.01, **: 0.05, *: 0.1}}\\
   \end{tabular}
\end{table}

\section{Receiver Operating Characteristic Curves for the Product Entry Model}\label{app_combined_rocs}

% Reset counters (to get A.1 instead of A.3)
\setcounter{table}{0} 
\setcounter{figure}{0} 
\setcounter{equation}{0} 

\begin{figure}[H]
    \centering
    \includegraphics[width=0.85\linewidth]{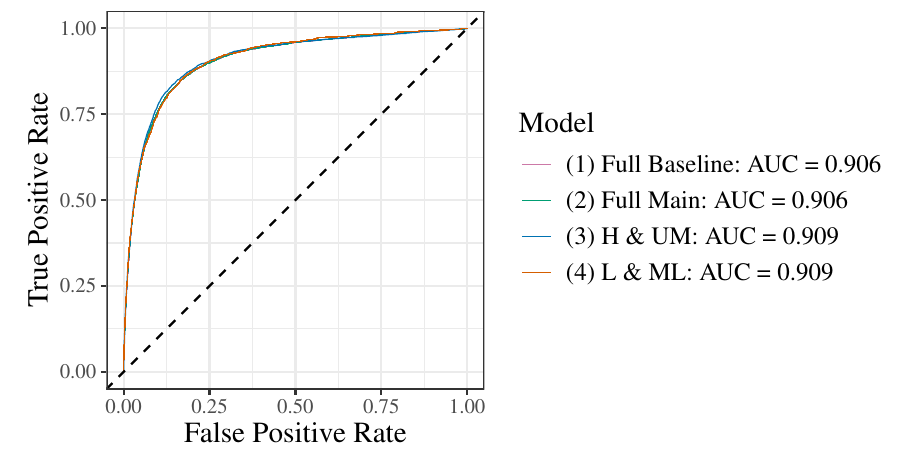}
    \caption{ROC Curves and AUC Values for Product Entry Models}
    \label{fig_combined_rocs}
\end{figure}

Fig.~\ref{fig_combined_rocs} displays the in-sample receiver operating characteristic (ROC) curves for product entry models (Model B): (1) full sample without interaction terms, (2) full sample with interaction terms, (3) high- and upper-middle-income countries with interaction terms, (4) lower-middle- and low-income countries with interaction terms (see Table~\ref{modB}). 

\section{Variance Inflation Factors of the Product Entry Model}\label{app_vifs}

% Reset counters (to get A.1 instead of A.3)
\setcounter{table}{0} 
\setcounter{figure}{0} 
\setcounter{equation}{0} 

\vspace{-10pt}
\noindent
\begin{minipage}[t]{0.48\textwidth}
  \centering
  \label{tab:vif_orth}
  \begin{table}[H]

\caption{VIFs for Product Entry Model (with orthogonalisation)}
\small
\begin{tabular}[t]{lr}
\toprule
Variable & VIF\\
\midrule
$\omega_{cpt-1}$ & 1.25\\
$\log(\mathit{RCA}_{cpt-1})$ & 1.27\\
$\mathit{PCI}_{pt-1}$ & 3.90\\
$\mathit{PSPI}_{pt-1}^{\perp}$ & 7.96\\
$\mathit{PEPI}_{pt-1}^{\perp}$ & 7.48\\
\addlinespace
$\omega_{cpt-1} \times \mathit{PCI}_{pt-1}$ & 3.80\\
$\omega_{cpt-1} \times \mathit{PSPI}_{pt-1}^{\perp}$ & 8.43\\
$\omega_{cpt-1} \times \mathit{PEPI}_{pt-1}^{\perp}$ & 7.98\\
\bottomrule
\end{tabular}
\end{table}
\end{minipage}%
\hfill
\begin{minipage}[t]{0.48\textwidth}
  \centering
  \label{tab:vif_notorth}
  \begin{table}[H]

\caption{VIFs for Product Entry Model (without orthogonalisation)}
\small
\begin{tabular}[t]{lr}
\toprule
Variable & VIF\\
\midrule
$\omega_{cpt-1}$ & 1.25\\
$\log(\mathit{RCA}_{cpt-1})$ & 1.27\\
$\mathit{PCI}_{pt-1}$ & 13.86\\
$\mathit{PSPI}_{pt-1}$ & 28.43\\
$\mathit{PEPI}_{pt-1}$ & 21.36\\
\addlinespace
$\omega_{cpt-1} \times \mathit{PCI}_{pt-1}$ & 14.02\\
$\omega_{cpt-1} \times \mathit{PSPI}_{pt-1}$ & 30.57\\
$\omega_{cpt-1} \times \mathit{PEPI}_{pt-1}$ & 24.05\\
\bottomrule
\end{tabular}
\end{table}
\end{minipage}

% \bibliographystyle{elsarticle-harv.bst} 
\bibliographystyle{elsarticle-num-names.bst} 
\bibliography{references.bib}

\begin{thebibliography}{107}
\expandafter\ifx\csname natexlab\endcsname\relax\def\natexlab#1{#1}\fi
\providecommand{\url}[1]{\texttt{#1}}
\providecommand{\href}[2]{#2}
\providecommand{\path}[1]{#1}
\providecommand{\DOIprefix}{doi:}
\providecommand{\ArXivprefix}{arXiv:}
\providecommand{\URLprefix}{URL: }
\providecommand{\Pubmedprefix}{pmid:}
\providecommand{\doi}[1]{\href{http://dx.doi.org/#1}{\path{#1}}}
\providecommand{\Pubmed}[1]{\href{pmid:#1}{\path{#1}}}
\providecommand{\bibinfo}[2]{#2}
\ifx\xfnm\relax \def\xfnm[#1]{\unskip,\space#1}\fi
%Type = Techreport
\bibitem[{Perrings(1994)}]{Perrings1994}
\bibinfo{author}{C.~Perrings}, \bibinfo{title}{Ecological {Resilience} in the
  {Sustainability} of {Economic} {Development}}, \bibinfo{type}{Technical
  Report} \bibinfo{number}{9405}, University of New York, \bibinfo{year}{1994}.
  \URLprefix \url{https://ageconsearch.umn.edu/record/263925}.
%Type = Article
\bibitem[{Nogueira et~al.(2023)Nogueira, Gomes, and Lopes}]{Nogueira.etal2023}
\bibinfo{author}{E.~Nogueira}, \bibinfo{author}{S.~Gomes},
  \bibinfo{author}{J.~M. Lopes},
\newblock \bibinfo{title}{Triple {Bottom} {Line}, {Sustainability}, and
  {Economic} {Development}: {What} {Binds} {Them} {Together}? {A}
  {Bibliometric} {Approach}},
\newblock \bibinfo{journal}{Sustainability} \bibinfo{volume}{15}
  (\bibinfo{year}{2023}) \bibinfo{pages}{6706}. \URLprefix
  \url{https://www.mdpi.com/2071-1050/15/8/6706}.
  \DOIprefix\doi{10.3390/su15086706}.
%Type = Article
\bibitem[{Boleti et~al.(2021)Boleti, Garas, Kyriakou, and
  Lapatinas}]{Boleti.etal2021}
\bibinfo{author}{E.~Boleti}, \bibinfo{author}{A.~Garas},
  \bibinfo{author}{A.~Kyriakou}, \bibinfo{author}{A.~Lapatinas},
\newblock \bibinfo{title}{Economic {Complexity} and {Environmental}
  {Performance}: {Evidence} from a {World} {Sample}},
\newblock \bibinfo{journal}{Environmental Modeling \& Assessment}
  \bibinfo{volume}{26} (\bibinfo{year}{2021}) \bibinfo{pages}{251--270}.
  \URLprefix \url{https://link.springer.com/10.1007/s10666-021-09750-0}.
  \DOIprefix\doi{10.1007/s10666-021-09750-0}.
%Type = Article
\bibitem[{Hidalgo et~al.(2007)Hidalgo, Klinger, Barabási, and
  Hausmann}]{Hidalgo.etal2007}
\bibinfo{author}{C.~A. Hidalgo}, \bibinfo{author}{B.~Klinger},
  \bibinfo{author}{A.-L. Barabási}, \bibinfo{author}{R.~Hausmann},
\newblock \bibinfo{title}{The {Product} {Space} {Conditions} the {Development}
  of {Nations}},
\newblock \bibinfo{journal}{Science} \bibinfo{volume}{317}
  (\bibinfo{year}{2007}) \bibinfo{pages}{482--487}. \URLprefix
  \url{https://www.science.org/doi/10.1126/science.1144581}.
  \DOIprefix\doi{10.1126/science.1144581}.
%Type = Article
\bibitem[{Hidalgo and Hausmann(2009)}]{Hidalgo.Hausmann2009}
\bibinfo{author}{C.~A. Hidalgo}, \bibinfo{author}{R.~Hausmann},
\newblock \bibinfo{title}{The building blocks of economic complexity},
\newblock \bibinfo{journal}{Proceedings of the National Academy of Sciences}
  \bibinfo{volume}{106} (\bibinfo{year}{2009}) \bibinfo{pages}{10570--10575}.
  \URLprefix \url{https://pnas.org/doi/full/10.1073/pnas.0900943106}.
  \DOIprefix\doi{10.1073/pnas.0900943106}.
%Type = Article
\bibitem[{Hausmann and Hidalgo(2011)}]{Hausmann.Hidalgo2011}
\bibinfo{author}{R.~Hausmann}, \bibinfo{author}{C.~A. Hidalgo},
\newblock \bibinfo{title}{The network structure of economic output},
\newblock \bibinfo{journal}{Journal of Economic Growth} \bibinfo{volume}{16}
  (\bibinfo{year}{2011}) \bibinfo{pages}{309--342}. \URLprefix
  \url{http://link.springer.com/10.1007/s10887-011-9071-4}.
  \DOIprefix\doi{10.1007/s10887-011-9071-4}.
%Type = Article
\bibitem[{Hidalgo(2021)}]{Hidalgo2021}
\bibinfo{author}{C.~A. Hidalgo},
\newblock \bibinfo{title}{Economic complexity theory and applications},
\newblock \bibinfo{journal}{Nature Reviews Physics} \bibinfo{volume}{3}
  (\bibinfo{year}{2021}) \bibinfo{pages}{92--113}. \URLprefix
  \url{https://www.nature.com/articles/s42254-020-00275-1}.
  \DOIprefix\doi{10.1038/s42254-020-00275-1}.
%Type = Article
\bibitem[{Pinheiro et~al.(2022)Pinheiro, Hartmann, Boschma, and
  Hidalgo}]{Pinheiro.etal2022}
\bibinfo{author}{F.~L. Pinheiro}, \bibinfo{author}{D.~Hartmann},
  \bibinfo{author}{R.~Boschma}, \bibinfo{author}{C.~A. Hidalgo},
\newblock \bibinfo{title}{The time and frequency of unrelated diversification},
\newblock \bibinfo{journal}{Research Policy} \bibinfo{volume}{51}
  (\bibinfo{year}{2022}) \bibinfo{pages}{104323}. \URLprefix
  \url{https://www.sciencedirect.com/science/article/pii/S0048733321001244}.
  \DOIprefix\doi{10.1016/j.respol.2021.104323}.
%Type = Article
\bibitem[{Caldarola et~al.(2024)Caldarola, Mazzilli, Napolitano, Patelli, and
  Sbardella}]{Caldarola.etal2024}
\bibinfo{author}{B.~Caldarola}, \bibinfo{author}{D.~Mazzilli},
  \bibinfo{author}{L.~Napolitano}, \bibinfo{author}{A.~Patelli},
  \bibinfo{author}{A.~Sbardella},
\newblock \bibinfo{title}{Economic complexity and the sustainability
  transition: a review of data, methods, and literature},
\newblock \bibinfo{journal}{Journal of Physics: Complexity} \bibinfo{volume}{5}
  (\bibinfo{year}{2024}) \bibinfo{pages}{022001}. \URLprefix
  \url{https://iopscience.iop.org/article/10.1088/2632-072X/ad4f3d}.
  \DOIprefix\doi{10.1088/2632-072X/ad4f3d}.
%Type = Article
\bibitem[{Montiel-Hernández et~al.(2024)Montiel-Hernández, Pérez-Hernández,
  and Salazar-Hernández}]{Montiel-Hernandez.etal2024}
\bibinfo{author}{M.~G. Montiel-Hernández}, \bibinfo{author}{C.~C.
  Pérez-Hernández}, \bibinfo{author}{B.~C. Salazar-Hernández},
\newblock \bibinfo{title}{The {Intrinsic} {Links} of {Economic} {Complexity}
  with {Sustainability} {Dimensions}: {A} {Systematic} {Review} and {Agenda}
  for {Future} {Research}},
\newblock \bibinfo{journal}{Sustainability} \bibinfo{volume}{16}
  (\bibinfo{year}{2024}) \bibinfo{pages}{391}. \URLprefix
  \url{https://www.mdpi.com/2071-1050/16/1/391}.
  \DOIprefix\doi{10.3390/su16010391}.
%Type = Article
\bibitem[{Ferraz et~al.(2021)Ferraz, Falguera, Mariano, and
  Hartmann}]{Ferraz.etal2021}
\bibinfo{author}{D.~Ferraz}, \bibinfo{author}{F.~P.~S. Falguera},
  \bibinfo{author}{E.~B. Mariano}, \bibinfo{author}{D.~Hartmann},
\newblock \bibinfo{title}{Linking {Economic} {Complexity}, {Diversification},
  and {Industrial} {Policy} with {Sustainable} {Development}: {A} {Structured}
  {Literature} {Review}},
\newblock \bibinfo{journal}{Sustainability} \bibinfo{volume}{13}
  (\bibinfo{year}{2021}) \bibinfo{pages}{1265}. \URLprefix
  \url{https://www.mdpi.com/2071-1050/13/3/1265}.
  \DOIprefix\doi{10.3390/su13031265}.
%Type = Article
\bibitem[{Fraccascia et~al.(2018)Fraccascia, Giannoccaro, and
  Albino}]{Fraccascia.etal2018}
\bibinfo{author}{L.~Fraccascia}, \bibinfo{author}{I.~Giannoccaro},
  \bibinfo{author}{V.~Albino},
\newblock \bibinfo{title}{Green product development: {What} does the country
  product space imply?},
\newblock \bibinfo{journal}{Journal of Cleaner Production}
  \bibinfo{volume}{170} (\bibinfo{year}{2018}) \bibinfo{pages}{1076--1088}.
  \URLprefix
  \url{https://linkinghub.elsevier.com/retrieve/pii/S0959652617321911}.
  \DOIprefix\doi{10.1016/j.jclepro.2017.09.190}.
%Type = Article
\bibitem[{Neffke et~al.(2011)Neffke, Henning, and Boschma}]{Neffke.etal2011}
\bibinfo{author}{F.~Neffke}, \bibinfo{author}{M.~Henning},
  \bibinfo{author}{R.~Boschma},
\newblock \bibinfo{title}{How {Do} {Regions} {Diversify} over {Time}?
  {Industry} {Relatedness} and the {Development} of {New} {Growth} {Paths} in
  {Regions}: {ECONOMIC} {GEOGRAPHY}},
\newblock \bibinfo{journal}{Economic Geography} \bibinfo{volume}{87}
  (\bibinfo{year}{2011}) \bibinfo{pages}{237--265}. \URLprefix
  \url{https://onlinelibrary.wiley.com/doi/10.1111/j.1944-8287.2011.01121.x}.
  \DOIprefix\doi{10.1111/j.1944-8287.2011.01121.x}.
%Type = Article
\bibitem[{Kogler et~al.(2013)Kogler, Rigby, and Tucker}]{Kogler.etal2013}
\bibinfo{author}{D.~F. Kogler}, \bibinfo{author}{D.~L. Rigby},
  \bibinfo{author}{I.~Tucker},
\newblock \bibinfo{title}{Mapping {Knowledge} {Space} and {Technological}
  {Relatedness} in {US} {Cities}},
\newblock \bibinfo{journal}{European Planning Studies} \bibinfo{volume}{21}
  (\bibinfo{year}{2013}) \bibinfo{pages}{1374--1391}. \URLprefix
  \url{https://doi.org/10.1080/09654313.2012.755832}.
  \DOIprefix\doi{10.1080/09654313.2012.755832}.
%Type = Article
\bibitem[{Muneepeerakul et~al.(2013)Muneepeerakul, Lobo, Shutters,
  Goméz-Liévano, and Qubbaj}]{Muneepeerakul.etal2013}
\bibinfo{author}{R.~Muneepeerakul}, \bibinfo{author}{J.~Lobo},
  \bibinfo{author}{S.~T. Shutters}, \bibinfo{author}{A.~Goméz-Liévano},
  \bibinfo{author}{M.~R. Qubbaj},
\newblock \bibinfo{title}{Urban {Economies} and {Occupation} {Space}: {Can}
  {They} {Get} “{There}” from “{Here}”?},
\newblock \bibinfo{journal}{PLOS ONE} \bibinfo{volume}{8}
  (\bibinfo{year}{2013}) \bibinfo{pages}{e73676}. \URLprefix
  \url{https://journals.plos.org/plosone/article?id=10.1371/journal.pone.0073676}.
  \DOIprefix\doi{10.1371/journal.pone.0073676}.
%Type = Article
\bibitem[{Neffke and Henning(2013)}]{Neffke.Henning2013}
\bibinfo{author}{F.~Neffke}, \bibinfo{author}{M.~Henning},
\newblock \bibinfo{title}{Skill relatedness and firm diversification},
\newblock \bibinfo{journal}{Strategic Management Journal} \bibinfo{volume}{34}
  (\bibinfo{year}{2013}) \bibinfo{pages}{297--316}. \URLprefix
  \url{https://onlinelibrary.wiley.com/doi/abs/10.1002/smj.2014}.
  \DOIprefix\doi{10.1002/smj.2014}.
%Type = Article
\bibitem[{Boschma et~al.(2013)Boschma, Minondo, and Navarro}]{Boschma.etal2013}
\bibinfo{author}{R.~Boschma}, \bibinfo{author}{A.~Minondo},
  \bibinfo{author}{M.~Navarro},
\newblock \bibinfo{title}{The {Emergence} of {New} {Industries} at the
  {Regional} {Level} in {Spain}: {A} {Proximity} {Approach} {Based} on
  {Product} {Relatedness}},
\newblock \bibinfo{journal}{Economic Geography} \bibinfo{volume}{89}
  (\bibinfo{year}{2013}) \bibinfo{pages}{29--51}. \URLprefix
  \url{https://onlinelibrary.wiley.com/doi/abs/10.1111/j.1944-8287.2012.01170.x}.
  \DOIprefix\doi{10.1111/j.1944-8287.2012.01170.x}.
%Type = Article
\bibitem[{Boschma et~al.(2015)Boschma, Balland, and Kogler}]{Boschma.etal2015}
\bibinfo{author}{R.~Boschma}, \bibinfo{author}{P.-A. Balland},
  \bibinfo{author}{D.~F. Kogler},
\newblock \bibinfo{title}{Relatedness and technological change in cities: the
  rise and fall of technological knowledge in {US} metropolitan areas from 1981
  to 2010},
\newblock \bibinfo{journal}{Industrial and Corporate Change}
  \bibinfo{volume}{24} (\bibinfo{year}{2015}) \bibinfo{pages}{223--250}.
  \URLprefix \url{https://doi.org/10.1093/icc/dtu012}.
  \DOIprefix\doi{10.1093/icc/dtu012}.
%Type = Article
\bibitem[{Poncet and De~Waldemar(2015)}]{Poncet.DeWaldemar2015}
\bibinfo{author}{S.~Poncet}, \bibinfo{author}{F.~S. De~Waldemar},
\newblock \bibinfo{title}{Product {Relatedness} and {Firm} {Exports} in
  {China}},
\newblock \bibinfo{journal}{The World Bank Economic Review}
  \bibinfo{volume}{29} (\bibinfo{year}{2015}) \bibinfo{pages}{579--605}.
  \URLprefix
  \url{https://academic.oup.com/wber/article-lookup/doi/10.1093/wber/lht037}.
  \DOIprefix\doi{10.1093/wber/lht037}.
%Type = Article
\bibitem[{Rigby(2015)}]{Rigby2015}
\bibinfo{author}{D.~L. Rigby},
\newblock \bibinfo{title}{Technological {Relatedness} and {Knowledge} {Space}:
  {Entry} and {Exit} of {US} {Cities} from {Patent} {Classes}},
\newblock \bibinfo{journal}{Regional Studies} \bibinfo{volume}{49}
  (\bibinfo{year}{2015}) \bibinfo{pages}{1922--1937}. \URLprefix
  \url{https://ideas.repec.org//a/taf/regstd/v49y2015i11p1922-1937.html}.
%Type = Article
\bibitem[{Guevara et~al.(2016)Guevara, Hartmann, Aristarán, Mendoza, and
  Hidalgo}]{Guevara.etal2016}
\bibinfo{author}{M.~R. Guevara}, \bibinfo{author}{D.~Hartmann},
  \bibinfo{author}{M.~Aristarán}, \bibinfo{author}{M.~Mendoza},
  \bibinfo{author}{C.~A. Hidalgo},
\newblock \bibinfo{title}{The research space: using career paths to predict the
  evolution of the research output of individuals, institutions, and nations},
\newblock \bibinfo{journal}{Scientometrics} \bibinfo{volume}{109}
  (\bibinfo{year}{2016}) \bibinfo{pages}{1695--1709}. \URLprefix
  \url{https://ideas.repec.org//a/spr/scient/v109y2016i3d10.1007_s11192-016-2125-9.html}.
%Type = Article
\bibitem[{Borggren et~al.(2016)Borggren, Eriksson, and
  Lindgren}]{Borggren.etal2016}
\bibinfo{author}{J.~Borggren}, \bibinfo{author}{R.~H. Eriksson},
  \bibinfo{author}{U.~Lindgren},
\newblock \bibinfo{title}{Knowledge flows in high-impact firms: {How} does
  relatedness influence survival, acquisition and exit?},
\newblock \bibinfo{journal}{Journal of Economic Geography} \bibinfo{volume}{16}
  (\bibinfo{year}{2016}) \bibinfo{pages}{637--665}. \URLprefix
  \url{https://ideas.repec.org//a/oup/jecgeo/v16y2016i3p637-665..html}.
%Type = Article
\bibitem[{Chen et~al.(2017)Chen, Poncet, and Xiong}]{Chen.etal2017}
\bibinfo{author}{Z.~Chen}, \bibinfo{author}{S.~Poncet},
  \bibinfo{author}{R.~Xiong},
\newblock \bibinfo{title}{Inter-industry relatedness and industrial-policy
  efficiency: {Evidence} from {China}’s export processing zones},
\newblock \bibinfo{journal}{Journal of Comparative Economics}
  \bibinfo{volume}{45} (\bibinfo{year}{2017}) \bibinfo{pages}{809--826}.
  \URLprefix
  \url{https://www.sciencedirect.com/science/article/pii/S0147596716000044}.
  \DOIprefix\doi{10.1016/j.jce.2016.01.003}.
%Type = Article
\bibitem[{Jara-Figueroa et~al.(2018)Jara-Figueroa, Jun, Glaeser, and
  Hidalgo}]{Jara-Figueroa.etal2018}
\bibinfo{author}{C.~Jara-Figueroa}, \bibinfo{author}{B.~Jun},
  \bibinfo{author}{E.~L. Glaeser}, \bibinfo{author}{C.~A. Hidalgo},
\newblock \bibinfo{title}{The role of industry-specific, occupation-specific,
  and location-specific knowledge in the growth and survival of new firms},
\newblock \bibinfo{journal}{Proceedings of the National Academy of Sciences}
  \bibinfo{volume}{115} (\bibinfo{year}{2018}) \bibinfo{pages}{12646--12653}.
  \URLprefix \url{https://www.pnas.org/doi/10.1073/pnas.1800475115}.
  \DOIprefix\doi{10.1073/pnas.1800475115}.
%Type = Misc
\bibitem[{Pinheiro et~al.(2018)Pinheiro, Alshamsi, Hartmann, Boschma, and
  Hidalgo}]{Pinheiro.etal2018}
\bibinfo{author}{F.~L. Pinheiro}, \bibinfo{author}{A.~Alshamsi},
  \bibinfo{author}{D.~Hartmann}, \bibinfo{author}{R.~Boschma},
  \bibinfo{author}{C.~A. Hidalgo}, \bibinfo{title}{Shooting {High} or {Low}:
  {Do} {Countries} {Benefit} from {Entering} {Unrelated} {Activities}?},
  \bibinfo{year}{2018}. \URLprefix \url{https://arxiv.org/abs/1801.05352}.
  \DOIprefix\doi{10.48550/ARXIV.1801.05352}.
%Type = Article
\bibitem[{Alabdulkareem et~al.(2018)Alabdulkareem, Frank, Sun, AlShebli,
  Hidalgo, and Rahwan}]{Alabdulkareem.etal2018}
\bibinfo{author}{A.~Alabdulkareem}, \bibinfo{author}{M.~R. Frank},
  \bibinfo{author}{L.~Sun}, \bibinfo{author}{B.~AlShebli},
  \bibinfo{author}{C.~Hidalgo}, \bibinfo{author}{I.~Rahwan},
\newblock \bibinfo{title}{Unpacking the polarization of workplace skills},
\newblock \bibinfo{journal}{Science Advances} \bibinfo{volume}{4}
  (\bibinfo{year}{2018}) \bibinfo{pages}{eaao6030}. \URLprefix
  \url{https://www.science.org/doi/10.1126/sciadv.aao6030}.
  \DOIprefix\doi{10.1126/sciadv.aao6030}.
%Type = Article
\bibitem[{Farinha et~al.(2019)Farinha, Balland, Morrison, and
  Boschma}]{Farinha.etal2019}
\bibinfo{author}{T.~Farinha}, \bibinfo{author}{P.-A. Balland},
  \bibinfo{author}{A.~Morrison}, \bibinfo{author}{R.~Boschma},
\newblock \bibinfo{title}{What drives the geography of jobs in the {US}?
  {Unpacking} relatedness},
\newblock \bibinfo{journal}{Industry and Innovation} \bibinfo{volume}{26}
  (\bibinfo{year}{2019}) \bibinfo{pages}{988--1022}. \URLprefix
  \url{https://doi.org/10.1080/13662716.2019.1591940}.
  \DOIprefix\doi{10.1080/13662716.2019.1591940}.
%Type = Misc
\bibitem[{Hartmann et~al.(2019)Hartmann, Bezerra, and
  Pinheiro}]{Hartmann.etal2019}
\bibinfo{author}{D.~Hartmann}, \bibinfo{author}{M.~Bezerra},
  \bibinfo{author}{F.~L. Pinheiro}, \bibinfo{title}{Identifying {Smart}
  {Strategies} for {Economic} {Diversification} and {Inclusive} {Growth} in
  {Developing} {Economies}. {The} {Case} of {Paraguay}}, \bibinfo{year}{2019}.
  \URLprefix \url{https://papers.ssrn.com/abstract=3346790}.
  \DOIprefix\doi{10.2139/ssrn.3346790}.
%Type = Article
\bibitem[{Chinazzi et~al.(2019)Chinazzi, Gonçalves, Zhang, and
  Vespignani}]{Chinazzi.etal2019}
\bibinfo{author}{M.~Chinazzi}, \bibinfo{author}{B.~Gonçalves},
  \bibinfo{author}{Q.~Zhang}, \bibinfo{author}{A.~Vespignani},
\newblock \bibinfo{title}{Mapping the physics research space: a machine
  learning approach},
\newblock \bibinfo{journal}{EPJ Data Science} \bibinfo{volume}{8}
  (\bibinfo{year}{2019}) \bibinfo{pages}{1--18}. \URLprefix
  \url{https://epjdatascience.springeropen.com/articles/10.1140/epjds/s13688-019-0210-z}.
  \DOIprefix\doi{10.1140/epjds/s13688-019-0210-z}.
%Type = Article
\bibitem[{Cicerone et~al.(2020)Cicerone, McCann, and
  Venhorst}]{Cicerone.etal2020}
\bibinfo{author}{G.~Cicerone}, \bibinfo{author}{P.~McCann},
  \bibinfo{author}{V.~A. Venhorst},
\newblock \bibinfo{title}{Promoting regional growth and innovation:
  relatedness, revealed comparative advantage and the product space},
\newblock \bibinfo{journal}{Journal of Economic Geography} \bibinfo{volume}{20}
  (\bibinfo{year}{2020}) \bibinfo{pages}{293--316}. \URLprefix
  \url{https://doi.org/10.1093/jeg/lbz001}. \DOIprefix\doi{10.1093/jeg/lbz001}.
%Type = Article
\bibitem[{Juhász et~al.(2021)Juhász, Broekel, and Boschma}]{Juhasz.etal2021}
\bibinfo{author}{S.~Juhász}, \bibinfo{author}{T.~Broekel},
  \bibinfo{author}{R.~Boschma},
\newblock \bibinfo{title}{Explaining the dynamics of relatedness: {The} role of
  co-location and complexity},
\newblock \bibinfo{journal}{Papers in Regional Science} \bibinfo{volume}{100}
  (\bibinfo{year}{2021}) \bibinfo{pages}{3--21}. \URLprefix
  \url{https://onlinelibrary.wiley.com/doi/abs/10.1111/pirs.12567}.
  \DOIprefix\doi{10.1111/pirs.12567}.
%Type = Article
\bibitem[{Hartmann et~al.(2021)Hartmann, Zagato, Gala, and
  Pinheiro}]{Hartmann.etal2021}
\bibinfo{author}{D.~Hartmann}, \bibinfo{author}{L.~Zagato},
  \bibinfo{author}{P.~Gala}, \bibinfo{author}{F.~L. Pinheiro},
\newblock \bibinfo{title}{Why did some countries catch-up, while others got
  stuck in the middle? {Stages} of productive sophistication and smart
  industrial policies},
\newblock \bibinfo{journal}{Structural Change and Economic Dynamics}
  \bibinfo{volume}{58} (\bibinfo{year}{2021}) \bibinfo{pages}{1--13}.
  \URLprefix
  \url{https://www.sciencedirect.com/science/article/pii/S0954349X21000412}.
  \DOIprefix\doi{10.1016/j.strueco.2021.04.007}.
%Type = Article
\bibitem[{Bam et~al.(2021)Bam, De~Bruyne, and Laing}]{Bam.etal2021}
\bibinfo{author}{W.~Bam}, \bibinfo{author}{K.~De~Bruyne},
  \bibinfo{author}{M.~Laing},
\newblock \bibinfo{title}{The {IO}–{PS} in the context of {GVC}-related
  policymaking: {The} case of the {South} {African} automotive industry},
\newblock \bibinfo{journal}{Journal of International Business Policy}
  \bibinfo{volume}{4} (\bibinfo{year}{2021}) \bibinfo{pages}{410--432}.
  \URLprefix \url{https://link.springer.com/10.1057/s42214-020-00081-7}.
  \DOIprefix\doi{10.1057/s42214-020-00081-7}.
%Type = Article
\bibitem[{Bam and De~Bruyne(2022)}]{Bam.DeBruyne2022}
\bibinfo{author}{W.~Bam}, \bibinfo{author}{K.~De~Bruyne},
\newblock \bibinfo{title}{The product space, sustainability, and {GVC} oriented
  industrial policies: {The} case of iron and steel in the {SACU}},
\newblock \bibinfo{journal}{Africa Journal of Management} \bibinfo{volume}{8}
  (\bibinfo{year}{2022}) \bibinfo{pages}{15--35}. \URLprefix
  \url{https://www.tandfonline.com/doi/full/10.1080/23322373.2021.2001289}.
  \DOIprefix\doi{10.1080/23322373.2021.2001289}.
%Type = Article
\bibitem[{De~Bruyne et~al.(2023)De~Bruyne, Bam, and
  Engelbrecht}]{DeBruyne.etal2023}
\bibinfo{author}{K.~De~Bruyne}, \bibinfo{author}{W.~Bam},
  \bibinfo{author}{D.~Engelbrecht},
\newblock \bibinfo{title}{South {Africa}'s titanium industrial policy: {A}
  product space perspective},
\newblock \bibinfo{journal}{South African Journal of Economics}
  \bibinfo{volume}{91} (\bibinfo{year}{2023}) \bibinfo{pages}{3--27}.
  \URLprefix \url{https://onlinelibrary.wiley.com/doi/10.1111/saje.12333}.
  \DOIprefix\doi{10.1111/saje.12333}.
%Type = Article
\bibitem[{Stephany and Teutloff(2024)}]{Stephany.Teutloff2024}
\bibinfo{author}{F.~Stephany}, \bibinfo{author}{O.~Teutloff},
\newblock \bibinfo{title}{What is the price of a skill? {The} value of
  complementarity},
\newblock \bibinfo{journal}{Research Policy} \bibinfo{volume}{53}
  (\bibinfo{year}{2024}) \bibinfo{pages}{104898}. \URLprefix
  \url{https://www.sciencedirect.com/science/article/pii/S0048733323001828}.
  \DOIprefix\doi{10.1016/j.respol.2023.104898}.
%Type = Article
\bibitem[{Pinheiro(2024)}]{Pinheiro2024}
\bibinfo{author}{C.~Pinheiro},
\newblock \bibinfo{title}{Relatedness and economic complexity as tools for
  industrial policy: {Insights} and limitations},
\newblock \bibinfo{journal}{Structural Change and Economic Dynamics}
  (\bibinfo{year}{2024}). \URLprefix
  \url{https://www.sciencedirect.com/science/article/pii/S0954349X24001486}.
  \DOIprefix\doi{10.1016/j.strueco.2024.09.019}.
%Type = Article
\bibitem[{Timbe et~al.(2024)Timbe, Pinheiro, Bam, Hartmann, and
  De~Bruyne}]{Timbe.etal2024}
\bibinfo{author}{G.~Timbe}, \bibinfo{author}{F.~L. Pinheiro},
  \bibinfo{author}{W.~Bam}, \bibinfo{author}{D.~Hartmann},
  \bibinfo{author}{K.~De~Bruyne},
\newblock \bibinfo{title}{Is natural resource abundance a curse or an
  opportunity? {Economic} complexity, {FDI}, and industrial policies in
  {Mozambique}},
\newblock \bibinfo{journal}{Resources Policy} \bibinfo{volume}{98}
  (\bibinfo{year}{2024}) \bibinfo{pages}{105326}. \URLprefix
  \url{https://www.sciencedirect.com/science/article/pii/S0301420724006937}.
  \DOIprefix\doi{10.1016/j.resourpol.2024.105326}.
%Type = Article
\bibitem[{Pinheiro et~al.(2025)Pinheiro, Balland, Boschma, and
  Hartmann}]{Pinheiro.etal2025}
\bibinfo{author}{F.~L. Pinheiro}, \bibinfo{author}{P.-A. Balland},
  \bibinfo{author}{R.~Boschma}, \bibinfo{author}{D.~Hartmann},
\newblock \bibinfo{title}{The dark side of the geography of innovation:
  relatedness, complexity and regional inequality in {Europe}},
\newblock \bibinfo{journal}{Regional Studies} \bibinfo{volume}{59}
  (\bibinfo{year}{2025}) \bibinfo{pages}{2106362}. \URLprefix
  \url{https://www.tandfonline.com/doi/full/10.1080/00343404.2022.2106362}.
  \DOIprefix\doi{10.1080/00343404.2022.2106362}.
%Type = Article
\bibitem[{Hartmann et~al.(2017)Hartmann, Guevara, Jara-Figueroa, Aristarán,
  and Hidalgo}]{Hartmann.etal2017}
\bibinfo{author}{D.~Hartmann}, \bibinfo{author}{M.~R. Guevara},
  \bibinfo{author}{C.~Jara-Figueroa}, \bibinfo{author}{M.~Aristarán},
  \bibinfo{author}{C.~A. Hidalgo},
\newblock \bibinfo{title}{Linking {Economic} {Complexity}, {Institutions}, and
  {Income} {Inequality}},
\newblock \bibinfo{journal}{World Development} \bibinfo{volume}{93}
  (\bibinfo{year}{2017}) \bibinfo{pages}{75--93}. \URLprefix
  \url{https://linkinghub.elsevier.com/retrieve/pii/S0305750X15309876}.
  \DOIprefix\doi{10.1016/j.worlddev.2016.12.020}.
%Type = Article
\bibitem[{Romero and Gramkow(2021)}]{Romero.Gramkow2021}
\bibinfo{author}{J.~P. Romero}, \bibinfo{author}{C.~Gramkow},
\newblock \bibinfo{title}{Economic complexity and greenhouse gas emissions},
\newblock \bibinfo{journal}{World Development} \bibinfo{volume}{139}
  (\bibinfo{year}{2021}) \bibinfo{pages}{105317}. \URLprefix
  \url{https://linkinghub.elsevier.com/retrieve/pii/S0305750X20304447}.
  \DOIprefix\doi{10.1016/j.worlddev.2020.105317}.
%Type = Techreport
\bibitem[{Lapatinas et~al.(2019)Lapatinas, Garas, Boleti, and
  Kyriakou}]{Lapatinas.etal2019}
\bibinfo{author}{A.~Lapatinas}, \bibinfo{author}{A.~Garas},
  \bibinfo{author}{E.~Boleti}, \bibinfo{author}{A.~Kyriakou},
  \bibinfo{title}{Economic complexity and environmental performance: {Evidence}
  from a world sample}, \bibinfo{type}{{MPRA} {Paper}} \bibinfo{number}{92833},
  University Library of Munich, Germany, \bibinfo{year}{2019}. \URLprefix
  \url{https://ideas.repec.org/p/pra/mprapa/92833.html}.
%Type = Article
\bibitem[{Mealy and Teytelboym(2022)}]{Mealy.Teytelboym2022}
\bibinfo{author}{P.~Mealy}, \bibinfo{author}{A.~Teytelboym},
\newblock \bibinfo{title}{Economic complexity and the green economy},
\newblock \bibinfo{journal}{Research Policy} \bibinfo{volume}{51}
  (\bibinfo{year}{2022}) \bibinfo{pages}{103948}. \URLprefix
  \url{https://linkinghub.elsevier.com/retrieve/pii/S0048733320300287}.
  \DOIprefix\doi{10.1016/j.respol.2020.103948}.
%Type = Techreport
\bibitem[{Hidalgo and Stojkoski(2025)}]{Hidalgo.Stojkoski2025}
\bibinfo{author}{C.~A. Hidalgo}, \bibinfo{author}{V.~Stojkoski},
  \bibinfo{title}{The {Theory} of {Economic} {Complexity}},
  \bibinfo{type}{Working {Paper}} \bibinfo{number}{1648}, Toulouse School of
  Economics, \bibinfo{year}{2025}. \URLprefix
  \url{https://www.tse-fr.eu/sites/default/files/TSE/documents/doc/wp/2025/wp_tse_1648.pdf}.
%Type = Article
\bibitem[{Balland et~al.(2022)Balland, Broekel, Diodato, Giuliani, Hausmann,
  O'Clery, and Rigby}]{Balland.etal2022}
\bibinfo{author}{P.-A. Balland}, \bibinfo{author}{T.~Broekel},
  \bibinfo{author}{D.~Diodato}, \bibinfo{author}{E.~Giuliani},
  \bibinfo{author}{R.~Hausmann}, \bibinfo{author}{N.~O'Clery},
  \bibinfo{author}{D.~Rigby},
\newblock \bibinfo{title}{Reprint of {The} new paradigm of economic
  complexity},
\newblock \bibinfo{journal}{Research Policy} \bibinfo{volume}{51}
  (\bibinfo{year}{2022}) \bibinfo{pages}{104568}. \URLprefix
  \url{https://linkinghub.elsevier.com/retrieve/pii/S0048733322000919}.
  \DOIprefix\doi{10.1016/j.respol.2022.104568}.
%Type = Article
\bibitem[{Schetter et~al.(2024)Schetter, Diodato, Protzer, Neffke, and
  Hausmann}]{Schetter.etal2024}
\bibinfo{author}{U.~Schetter}, \bibinfo{author}{D.~Diodato},
  \bibinfo{author}{E.~Protzer}, \bibinfo{author}{F.~Neffke},
  \bibinfo{author}{R.~Hausmann},
\newblock \bibinfo{title}{From {Products} to {Capabilities}: {Constructing} a
  {Genotypic} {Product} {Space}},
\newblock \bibinfo{journal}{Papers in Evolutionary Economic Geography (PEEG)}
  (\bibinfo{year}{2024}). \URLprefix
  \url{https://ideas.repec.org//p/egu/wpaper/2419.html}.
%Type = Article
\bibitem[{Cristelli et~al.(2013)Cristelli, Gabrielli, Tacchella, Caldarelli,
  and Pietronero}]{Cristelli.etal2013}
\bibinfo{author}{M.~C.~A. Cristelli}, \bibinfo{author}{A.~Gabrielli},
  \bibinfo{author}{A.~Tacchella}, \bibinfo{author}{G.~Caldarelli},
  \bibinfo{author}{L.~Pietronero},
\newblock \bibinfo{title}{Measuring the {Intangibles}: {A} {Metrics} for the
  {Economic} {Complexity} of {Countries} and {Products}},
\newblock \bibinfo{journal}{PLOS ONE} \bibinfo{volume}{8}
  (\bibinfo{year}{2013}) \bibinfo{pages}{e70726}. \URLprefix
  \url{https://journals.plos.org/plosone/article?id=10.1371/journal.pone.0070726}.
  \DOIprefix\doi{10.1371/journal.pone.0070726}.
%Type = Misc
\bibitem[{Cristelli et~al.(2017)Cristelli, Tacchella, Cader, Roster, and
  Pietronero}]{Cristelli.etal2017}
\bibinfo{author}{M.~C.~A. Cristelli}, \bibinfo{author}{A.~Tacchella},
  \bibinfo{author}{M.~Z. Cader}, \bibinfo{author}{K.~I. Roster},
  \bibinfo{author}{L.~Pietronero}, \bibinfo{title}{On the {Predictability} of
  {Growth}}, \bibinfo{year}{2017}. \URLprefix
  \url{https://papers.ssrn.com/abstract=3006151}.
%Type = Inproceedings
\bibitem[{Hidalgo et~al.(2018)Hidalgo, Balland, Boschma, Delgado, Feldman,
  Frenken, Glaeser, He, Kogler, Morrison, Neffke, Rigby, Stern, Zheng, and
  Zhu}]{Hidalgo.etal2018}
\bibinfo{author}{C.~A. Hidalgo}, \bibinfo{author}{P.-A. Balland},
  \bibinfo{author}{R.~Boschma}, \bibinfo{author}{M.~Delgado},
  \bibinfo{author}{M.~Feldman}, \bibinfo{author}{K.~Frenken},
  \bibinfo{author}{E.~Glaeser}, \bibinfo{author}{C.~He}, \bibinfo{author}{D.~F.
  Kogler}, \bibinfo{author}{A.~Morrison}, \bibinfo{author}{F.~Neffke},
  \bibinfo{author}{D.~Rigby}, \bibinfo{author}{S.~Stern},
  \bibinfo{author}{S.~Zheng}, \bibinfo{author}{S.~Zhu},
\newblock \bibinfo{title}{The {Principle} of {Relatedness}},
\newblock in: \bibinfo{editor}{A.~J. Morales}, \bibinfo{editor}{C.~Gershenson},
  \bibinfo{editor}{D.~Braha}, \bibinfo{editor}{A.~A. Minai},
  \bibinfo{editor}{Y.~Bar-Yam} (Eds.), \bibinfo{booktitle}{Unifying {Themes} in
  {Complex} {Systems} {IX}}, Springer {Proceedings} in {Complexity},
  \bibinfo{publisher}{Springer International Publishing},
  \bibinfo{address}{Cham}, \bibinfo{year}{2018}, pp. \bibinfo{pages}{451--457}.
  \DOIprefix\doi{10.1007/978-3-319-96661-8_46}.
%Type = Article
\bibitem[{Boschma and Capone(2015)}]{Boschma.Capone2015}
\bibinfo{author}{R.~Boschma}, \bibinfo{author}{G.~Capone},
\newblock \bibinfo{title}{Institutions and diversification: {Related} versus
  unrelated diversification in a varieties of capitalism framework},
\newblock \bibinfo{journal}{Research Policy} \bibinfo{volume}{44}
  (\bibinfo{year}{2015}) \bibinfo{pages}{1902--1914}. \URLprefix
  \url{https://www.sciencedirect.com/science/article/pii/S0048733315001109}.
  \DOIprefix\doi{10.1016/j.respol.2015.06.013}.
%Type = Article
\bibitem[{Zhu et~al.(2017)Zhu, Hey, and Zhou}]{Zhu.etal2017}
\bibinfo{author}{S.~Zhu}, \bibinfo{author}{C.~Hey}, \bibinfo{author}{Y.~Zhou},
\newblock \bibinfo{title}{How to jump further and catch up? {Path}-breaking in
  an uneven industry space},
\newblock \bibinfo{journal}{Journal of Economic Geography} \bibinfo{volume}{17}
  (\bibinfo{year}{2017}) \bibinfo{pages}{521--545}. \URLprefix
  \url{https://www.jstor.org/stable/26501035}.
%Type = Article
\bibitem[{Tacchella et~al.(2023)Tacchella, Zaccaria, Miccheli, and
  Pietronero}]{Tacchella.etal2023}
\bibinfo{author}{A.~Tacchella}, \bibinfo{author}{A.~Zaccaria},
  \bibinfo{author}{M.~Miccheli}, \bibinfo{author}{L.~Pietronero},
\newblock \bibinfo{title}{Relatedness in the era of machine learning},
\newblock \bibinfo{journal}{Chaos, Solitons \& Fractals} \bibinfo{volume}{176}
  (\bibinfo{year}{2023}) \bibinfo{pages}{114071}. \URLprefix
  \url{https://linkinghub.elsevier.com/retrieve/pii/S0960077923009724}.
  \DOIprefix\doi{10.1016/j.chaos.2023.114071}.
%Type = Book
\bibitem[{Hausmann and Hidalgo(2013)}]{Hausmann.Hidalgo2013}
\bibinfo{editor}{R.~Hausmann}, \bibinfo{editor}{C.~A. Hidalgo} (Eds.),
  \bibinfo{title}{The atlas of economic complexity: mapping paths to
  prosperity}, \bibinfo{edition}{updated edition} ed., \bibinfo{publisher}{The
  MIT Press}, \bibinfo{address}{Cambridge, MA}, \bibinfo{year}{2013}.
%Type = Article
\bibitem[{Domini(2019)}]{Domini2019}
\bibinfo{author}{G.~Domini},
\newblock \bibinfo{title}{Patterns of specialisation and economic complexity
  through the lens of universal exhibitions, 1855-1900},
\newblock \bibinfo{journal}{LEM Papers Series}  (\bibinfo{year}{2019}).
  \URLprefix \url{https://ideas.repec.org//p/ssa/lemwps/2019-20.html}.
%Type = Article
\bibitem[{Koch(2021)}]{Koch2021}
\bibinfo{author}{P.~Koch},
\newblock \bibinfo{title}{Economic complexity and growth: {Can} value-added
  exports better explain the link?},
\newblock \bibinfo{journal}{Economics Letters} \bibinfo{volume}{198}
  (\bibinfo{year}{2021}). \URLprefix
  \url{https://econpapers.repec.org/article/eeeecolet/v_3a198_3ay_3a2021_3ai_3ac_3as0165176520304420.htm}.
%Type = Article
\bibitem[{Lo~Turco and Maggioni(2022)}]{LoTurco.Maggioni2022}
\bibinfo{author}{A.~Lo~Turco}, \bibinfo{author}{D.~Maggioni},
\newblock \bibinfo{title}{The knowledge and skill content of production
  complexity},
\newblock \bibinfo{journal}{Research Policy} \bibinfo{volume}{51}
  (\bibinfo{year}{2022}) \bibinfo{pages}{104059}. \URLprefix
  \url{https://www.sciencedirect.com/science/article/pii/S0048733320301372}.
  \DOIprefix\doi{10.1016/j.respol.2020.104059}.
%Type = Article
\bibitem[{Ourens(2013)}]{Ourens2013}
\bibinfo{author}{G.~Ourens},
\newblock \bibinfo{title}{Can the {Method} of {Reflections} help predict future
  growth?},
\newblock \bibinfo{journal}{LIDAM Discussion Papers IRES}
  (\bibinfo{year}{2013}). \URLprefix
  \url{https://ideas.repec.org//p/ctl/louvir/2013008.html}.
%Type = Article
\bibitem[{Stojkoski et~al.(2016)Stojkoski, Utkovski, and
  Kocarev}]{Stojkoski.etal2016}
\bibinfo{author}{V.~Stojkoski}, \bibinfo{author}{Z.~Utkovski},
  \bibinfo{author}{L.~Kocarev},
\newblock \bibinfo{title}{The {Impact} of {Services} on {Economic}
  {Complexity}: {Service} {Sophistication} as {Route} for {Economic} {Growth}},
\newblock \bibinfo{journal}{PLoS ONE} \bibinfo{volume}{11}
  (\bibinfo{year}{2016}) \bibinfo{pages}{e0161633}. \URLprefix
  \url{https://www.ncbi.nlm.nih.gov/pmc/articles/PMC4999235/}.
  \DOIprefix\doi{10.1371/journal.pone.0161633}.
%Type = Article
\bibitem[{Felipe et~al.(2012)Felipe, Kumar, Abdon, and
  Bacate}]{Felipe.etal2012}
\bibinfo{author}{J.~Felipe}, \bibinfo{author}{U.~Kumar},
  \bibinfo{author}{A.~Abdon}, \bibinfo{author}{M.~Bacate},
\newblock \bibinfo{title}{Product complexity and economic development},
\newblock \bibinfo{journal}{Structural Change and Economic Dynamics}
  \bibinfo{volume}{23} (\bibinfo{year}{2012}) \bibinfo{pages}{36--68}.
  \URLprefix
  \url{https://www.sciencedirect.com/science/article/pii/S0954349X11000567}.
  \DOIprefix\doi{10.1016/j.strueco.2011.08.003}.
%Type = Article
\bibitem[{Stojkoski et~al.(2023)Stojkoski, Koch, and
  Hidalgo}]{Stojkoski.etal2023}
\bibinfo{author}{V.~Stojkoski}, \bibinfo{author}{P.~Koch},
  \bibinfo{author}{C.~A. Hidalgo},
\newblock \bibinfo{title}{Multidimensional economic complexity and inclusive
  green growth},
\newblock \bibinfo{journal}{Communications Earth \& Environment}
  \bibinfo{volume}{4} (\bibinfo{year}{2023}) \bibinfo{pages}{130}. \URLprefix
  \url{https://www.nature.com/articles/s43247-023-00770-0}.
  \DOIprefix\doi{10.1038/s43247-023-00770-0}.
%Type = Article
\bibitem[{Zhu and Li(2017)}]{Zhu.Li2017}
\bibinfo{author}{S.~Zhu}, \bibinfo{author}{R.~Li},
\newblock \bibinfo{title}{Economic complexity, human capital and economic
  growth: empirical research based on cross-country panel data},
\newblock \bibinfo{journal}{Applied Economics} \bibinfo{volume}{49}
  (\bibinfo{year}{2017}) \bibinfo{pages}{3815--3828}. \URLprefix
  \url{https://www.tandfonline.com/doi/full/10.1080/00036846.2016.1270413}.
  \DOIprefix\doi{10.1080/00036846.2016.1270413}.
%Type = Article
\bibitem[{Doğan et~al.(2021)Doğan, Driha, Balsalobre~Lorente, and
  Shahzad}]{Dogan.etal2021}
\bibinfo{author}{B.~Doğan}, \bibinfo{author}{O.~M. Driha},
  \bibinfo{author}{D.~Balsalobre~Lorente}, \bibinfo{author}{U.~Shahzad},
\newblock \bibinfo{title}{The mitigating effects of economic complexity and
  renewable energy on carbon emissions in developed countries},
\newblock \bibinfo{journal}{Sustainable Development} \bibinfo{volume}{29}
  (\bibinfo{year}{2021}) \bibinfo{pages}{1--12}. \URLprefix
  \url{https://onlinelibrary.wiley.com/doi/abs/10.1002/sd.2125}.
  \DOIprefix\doi{10.1002/sd.2125}.
%Type = Article
\bibitem[{Safi et~al.(2023)Safi, Wei, Sansaloni, and Umar}]{Safi.etal2023}
\bibinfo{author}{A.~Safi}, \bibinfo{author}{X.~Wei}, \bibinfo{author}{E.~M.
  Sansaloni}, \bibinfo{author}{M.~Umar},
\newblock \bibinfo{title}{Breaking down the complexity of sustainable
  development: {A} focus on resources, economic complexity, and innovation},
\newblock \bibinfo{journal}{Resources Policy} \bibinfo{volume}{83}
  (\bibinfo{year}{2023}). \URLprefix
  \url{https://ideas.repec.org//a/eee/jrpoli/v83y2023ics0301420723004579.html}.
%Type = Article
\bibitem[{Saqib and Dincă(2023)}]{Saqib.Dinca2023}
\bibinfo{author}{N.~Saqib}, \bibinfo{author}{G.~Dincă},
\newblock \bibinfo{title}{Exploring the asymmetric impact of economic
  complexity, {FDI}, and green technology on carbon emissions: {Policy}
  stringency for clean-energy investing countries},
\newblock \bibinfo{journal}{Geoscience Frontiers}  (\bibinfo{year}{2023})
  \bibinfo{pages}{101671}. \URLprefix
  \url{https://www.sciencedirect.com/science/article/pii/S167498712300138X}.
  \DOIprefix\doi{10.1016/j.gsf.2023.101671}.
%Type = Article
\bibitem[{Payne et~al.(2023)Payne, Truong, Chu, Doğan, and
  Ghosh}]{Payne.etal2023}
\bibinfo{author}{J.~E. Payne}, \bibinfo{author}{H.~H.~D. Truong},
  \bibinfo{author}{L.~K. Chu}, \bibinfo{author}{B.~Doğan},
  \bibinfo{author}{S.~Ghosh},
\newblock \bibinfo{title}{The effect of economic complexity and energy security
  on measures of energy efficiency: {Evidence} from panel quantile analysis},
\newblock \bibinfo{journal}{Energy Policy} \bibinfo{volume}{177}
  (\bibinfo{year}{2023}). \URLprefix
  \url{https://ideas.repec.org//a/eee/enepol/v177y2023ics0301421523001325.html}.
%Type = Article
\bibitem[{Balsalobre~Lorente et~al.(2023)Balsalobre~Lorente, Joof, Samour, and
  Türsoy}]{BalsalobreLorente.etal2023}
\bibinfo{author}{D.~Balsalobre~Lorente}, \bibinfo{author}{F.~Joof},
  \bibinfo{author}{A.~Samour}, \bibinfo{author}{T.~Türsoy},
\newblock \bibinfo{title}{Renewable energy, economic complexity and
  biodiversity risk: {New} insights from {China}},
\newblock \bibinfo{journal}{Environmental and Sustainability Indicators}
  \bibinfo{volume}{18} (\bibinfo{year}{2023}) \bibinfo{pages}{100244}.
  \URLprefix
  \url{https://www.sciencedirect.com/science/article/pii/S2665972723000211}.
  \DOIprefix\doi{10.1016/j.indic.2023.100244}.
%Type = Article
\bibitem[{Rafique et~al.(2022)Rafique, Nadeem, Xia, Ikram, Shoaib, and
  Shahzad}]{Rafique.etal2022}
\bibinfo{author}{M.~Z. Rafique}, \bibinfo{author}{A.~M. Nadeem},
  \bibinfo{author}{W.~Xia}, \bibinfo{author}{M.~Ikram}, \bibinfo{author}{H.~M.
  Shoaib}, \bibinfo{author}{U.~Shahzad},
\newblock \bibinfo{title}{Does economic complexity matter for environmental
  sustainability? {Using} ecological footprint as an indicator},
\newblock \bibinfo{journal}{Environment, Development and Sustainability}
  \bibinfo{volume}{24} (\bibinfo{year}{2022}) \bibinfo{pages}{4623--4640}.
  \DOIprefix\doi{10.1007/s10668-021-01625-4}.
%Type = Misc
\bibitem[{Grossman and Krueger(1991)}]{Grossman.Krueger1991}
\bibinfo{author}{G.~M. Grossman}, \bibinfo{author}{A.~B. Krueger},
  \bibinfo{title}{Environmental {Impacts} of a {North} {American} {Free}
  {Trade} {Agreement}}, \bibinfo{year}{1991}. \URLprefix
  \url{https://www.nber.org/papers/w3914}. \DOIprefix\doi{10.3386/w3914}.
%Type = Article
\bibitem[{Grossman and Krueger(1995)}]{Grossman.Krueger1995}
\bibinfo{author}{G.~M. Grossman}, \bibinfo{author}{A.~Krueger},
\newblock \bibinfo{title}{Economic {Growth} and the {Environment}},
\newblock \bibinfo{journal}{The Quarterly Journal of Economics}
  \bibinfo{volume}{110} (\bibinfo{year}{1995}) \bibinfo{pages}{353--377}.
  \URLprefix
  \url{https://econpapers.repec.org/article/oupqjecon/v_3a110_3ay_3a1995_3ai_3a2_3ap_3a353-377..htm}.
%Type = Article
\bibitem[{Kiliç et~al.(2024)Kiliç, Soyyiğit, and Bayrakdar}]{Kilic.etal2024}
\bibinfo{author}{C.~Kiliç}, \bibinfo{author}{S.~Soyyiğit},
  \bibinfo{author}{S.~Bayrakdar},
\newblock \bibinfo{title}{Economic {Complexity}, {Ecological} {Footprint}, and
  the {Environmental} {Kuznets} {Curve}: {Findings} from {Selected}
  {Industrialized} {Countries}},
\newblock \bibinfo{journal}{Journal of the Knowledge Economy}
  \bibinfo{volume}{15} (\bibinfo{year}{2024}) \bibinfo{pages}{7402--7427}.
  \URLprefix \url{https://doi.org/10.1007/s13132-023-01411-9}.
  \DOIprefix\doi{10.1007/s13132-023-01411-9}.
%Type = Article
\bibitem[{Neagu(2019)}]{Neagu2019}
\bibinfo{author}{O.~Neagu},
\newblock \bibinfo{title}{The {Link} between {Economic} {Complexity} and
  {Carbon} {Emissions} in the {European} {Union} {Countries}: {A} {Model}
  {Based} on the {Environmental} {Kuznets} {Curve} ({EKC}) {Approach}},
\newblock \bibinfo{journal}{Sustainability} \bibinfo{volume}{11}
  (\bibinfo{year}{2019}) \bibinfo{pages}{4753}. \URLprefix
  \url{https://www.mdpi.com/2071-1050/11/17/4753}.
  \DOIprefix\doi{10.3390/su11174753}.
%Type = Article
\bibitem[{Doğan et~al.(2022)Doğan, Ghosh, Shahzadi, Balsalobre-Lorente, and
  Nguyen}]{Dogan.etal2022}
\bibinfo{author}{B.~Doğan}, \bibinfo{author}{S.~Ghosh},
  \bibinfo{author}{I.~Shahzadi}, \bibinfo{author}{D.~Balsalobre-Lorente},
  \bibinfo{author}{C.~P. Nguyen},
\newblock \bibinfo{title}{The relevance of economic complexity and economic
  globalization as determinants of energy demand for different stages of
  development},
\newblock \bibinfo{journal}{Renewable Energy} \bibinfo{volume}{190}
  (\bibinfo{year}{2022}) \bibinfo{pages}{371--384}. \URLprefix
  \url{https://linkinghub.elsevier.com/retrieve/pii/S0960148122004013}.
  \DOIprefix\doi{10.1016/j.renene.2022.03.117}.
%Type = Article
\bibitem[{Fawaz and Rahnama-Moghadamm(2019)}]{Fawaz.Rahnama-Moghadamm2019}
\bibinfo{author}{F.~Fawaz}, \bibinfo{author}{M.~Rahnama-Moghadamm},
\newblock \bibinfo{title}{Spatial dependence of global income inequality: {The}
  role of economic complexity},
\newblock \bibinfo{journal}{The International Trade Journal}
  \bibinfo{volume}{33} (\bibinfo{year}{2019}) \bibinfo{pages}{542--554}.
  \URLprefix
  \url{https://www.tandfonline.com/doi/full/10.1080/08853908.2018.1535336}.
  \DOIprefix\doi{10.1080/08853908.2018.1535336}.
%Type = Article
\bibitem[{Vu(2020)}]{Vu2020a}
\bibinfo{author}{T.~V. Vu},
\newblock \bibinfo{title}{Does {LGBT} {Inclusion} {Promote} {National}
  {Innovative} {Capacity}?},
\newblock \bibinfo{journal}{SSRN Electronic Journal}  (\bibinfo{year}{2020}).
  \URLprefix \url{https://www.ssrn.com/abstract=3523553}.
  \DOIprefix\doi{10.2139/ssrn.3523553}.
%Type = Article
\bibitem[{Vu(2022)}]{Vu2022a}
\bibinfo{author}{T.~V. Vu},
\newblock \bibinfo{title}{Linking {LGBT} inclusion and national innovative
  capacity},
\newblock \bibinfo{journal}{Social Indicators Research} \bibinfo{volume}{159}
  (\bibinfo{year}{2022}) \bibinfo{pages}{191--214}. \URLprefix
  \url{https://doi.org/10.1007/s11205-021-02743-2}.
  \DOIprefix\doi{10.1007/s11205-021-02743-2}.
%Type = Article
\bibitem[{Vu(2020)}]{Vu2020}
\bibinfo{author}{T.~V. Vu},
\newblock \bibinfo{title}{Economic complexity and health outcomes: {A} global
  perspective},
\newblock \bibinfo{journal}{Social Science \& Medicine} \bibinfo{volume}{265}
  (\bibinfo{year}{2020}) \bibinfo{pages}{113480}. \URLprefix
  \url{https://www.sciencedirect.com/science/article/pii/S0277953620306997}.
  \DOIprefix\doi{10.1016/j.socscimed.2020.113480}.
%Type = Article
\bibitem[{Ben~Saad and Assoumou-Ella(2019)}]{BenSaad.Assoumou-Ella2019}
\bibinfo{author}{M.~Ben~Saad}, \bibinfo{author}{G.~Assoumou-Ella},
\newblock \bibinfo{title}{Economic {Complexity} and {Gender} {Inequality} in
  {Education}: {An} {Empirical} {Study}},
\newblock \bibinfo{journal}{SSRN Electronic Journal}  (\bibinfo{year}{2019}).
  \URLprefix \url{https://www.ssrn.com/abstract=3340913}.
  \DOIprefix\doi{10.2139/ssrn.3340913}.
%Type = Misc
\bibitem[{Barza et~al.(2020)Barza, Jara-Figueroa, Hidalgo, and
  Viarengo}]{Barza.etal2020}
\bibinfo{author}{R.~Barza}, \bibinfo{author}{C.~Jara-Figueroa},
  \bibinfo{author}{C.~A. Hidalgo}, \bibinfo{author}{M.~Viarengo},
  \bibinfo{title}{Knowledge {Intensity} and {Gender} {Wage} {Gaps}: {Evidence}
  from {Linked} {Employer}-{Employee} {Data}}, \bibinfo{year}{2020}. \URLprefix
  \url{https://papers.ssrn.com/abstract=3689464}.
  \DOIprefix\doi{10.2139/ssrn.3689464}.
%Type = Article
\bibitem[{Adam et~al.(2023)Adam, Garas, Katsaiti, and
  Lapatinas}]{Adam.etal2023}
\bibinfo{author}{A.~Adam}, \bibinfo{author}{A.~Garas}, \bibinfo{author}{M.-S.
  Katsaiti}, \bibinfo{author}{A.~Lapatinas},
\newblock \bibinfo{title}{Economic complexity and jobs: an empirical analysis},
\newblock \bibinfo{journal}{Economics of Innovation and New Technology}
  \bibinfo{volume}{32} (\bibinfo{year}{2023}) \bibinfo{pages}{25--52}.
  \URLprefix \url{https://doi.org/10.1080/10438599.2020.1859751}.
  \DOIprefix\doi{10.1080/10438599.2020.1859751}.
%Type = Article
\bibitem[{Nguyen et~al.(2023)Nguyen, Nguyen, and Tran}]{Nguyen.etal2023}
\bibinfo{author}{C.~P. Nguyen}, \bibinfo{author}{B.~Q. Nguyen},
  \bibinfo{author}{D.~T.~L. Tran},
\newblock \bibinfo{title}{Economic complexity and income inequality: {New}
  evidence of a nonlinear effect},
\newblock \bibinfo{journal}{Social Science Quarterly} \bibinfo{volume}{104}
  (\bibinfo{year}{2023}) \bibinfo{pages}{829--868}. \URLprefix
  \url{https://ideas.repec.org//a/bla/socsci/v104y2023i4p829-868.html}.
%Type = Article
\bibitem[{Vu(2022)}]{Vu2022}
\bibinfo{author}{T.~V. Vu},
\newblock \bibinfo{title}{Does institutional quality foster economic
  complexity? {The} fundamental drivers of productive capabilities},
\newblock \bibinfo{journal}{Empirical Economics} \bibinfo{volume}{63}
  (\bibinfo{year}{2022}) \bibinfo{pages}{1571--1604}. \URLprefix
  \url{https://doi.org/10.1007/s00181-021-02175-4}.
  \DOIprefix\doi{10.1007/s00181-021-02175-4}.
%Type = Techreport
\bibitem[{Morais et~al.(2018)Morais, Swart, and Jordaan}]{Morais.etal2018}
\bibinfo{author}{M.~B. Morais}, \bibinfo{author}{J.~Swart},
  \bibinfo{author}{J.~A. Jordaan}, \bibinfo{title}{Economic {Complexity} and
  {Inequality}: {Does} {Productive} {Structure} {Affect} {Regional} {Wage}
  {Differentials} in {Brazil}?}, \bibinfo{type}{Technical Report}, Utrecht
  School of Economics, \bibinfo{year}{2018}. \URLprefix
  \url{https://ideas.repec.org//p/use/tkiwps/1811.html}.
%Type = Article
\bibitem[{Sbardella et~al.(2017)Sbardella, Pugliese, and
  Pietronero}]{Sbardella.etal2017}
\bibinfo{author}{A.~Sbardella}, \bibinfo{author}{E.~Pugliese},
  \bibinfo{author}{L.~Pietronero},
\newblock \bibinfo{title}{Economic development and wage inequality: {A} complex
  system analysis},
\newblock \bibinfo{journal}{PLOS ONE} \bibinfo{volume}{12}
  (\bibinfo{year}{2017}) \bibinfo{pages}{e0182774}. \URLprefix
  \url{https://dx.plos.org/10.1371/journal.pone.0182774}.
  \DOIprefix\doi{10.1371/journal.pone.0182774}.
%Type = Article
\bibitem[{Ghosh et~al.(2023)Ghosh, Doğan, Can, Shah, and
  Apergis}]{Ghosh.etal2023}
\bibinfo{author}{S.~Ghosh}, \bibinfo{author}{B.~Doğan},
  \bibinfo{author}{M.~Can}, \bibinfo{author}{M.~I. Shah},
  \bibinfo{author}{N.~Apergis},
\newblock \bibinfo{title}{Does economic structure matter for income
  inequality?},
\newblock \bibinfo{journal}{Quality \& Quantity} \bibinfo{volume}{57}
  (\bibinfo{year}{2023}) \bibinfo{pages}{2507--2527}. \URLprefix
  \url{https://doi.org/10.1007/s11135-022-01462-1}.
  \DOIprefix\doi{10.1007/s11135-022-01462-1}.
%Type = Article
\bibitem[{Le~Caous and Huarng(2020)}]{LeCaous.Huarng2020}
\bibinfo{author}{E.~Le~Caous}, \bibinfo{author}{F.~Huarng},
\newblock \bibinfo{title}{Economic {Complexity} and the {Mediating} {Effects}
  of {Income} {Inequality}: {Reaching} {Sustainable} {Development} in
  {Developing} {Countries}},
\newblock \bibinfo{journal}{Sustainability} \bibinfo{volume}{12}
  (\bibinfo{year}{2020}) \bibinfo{pages}{2089}. \URLprefix
  \url{https://www.mdpi.com/2071-1050/12/5/2089}.
  \DOIprefix\doi{10.3390/su12052089}.
%Type = Article
\bibitem[{Djeunankan et~al.(2025)Djeunankan, Tadadjeu, Njangang, and
  Mazhar}]{Djeunankan.etal2025}
\bibinfo{author}{R.~Djeunankan}, \bibinfo{author}{S.~Tadadjeu},
  \bibinfo{author}{H.~Njangang}, \bibinfo{author}{U.~Mazhar},
\newblock \bibinfo{title}{The hidden cost of sophistication: economic
  complexity and obesity},
\newblock \bibinfo{journal}{The European journal of health economics: HEPAC:
  health economics in prevention and care} \bibinfo{volume}{26}
  (\bibinfo{year}{2025}) \bibinfo{pages}{243--265}.
  \DOIprefix\doi{10.1007/s10198-024-01699-7}.
%Type = Article
\bibitem[{Hausmann et~al.(2007)Hausmann, Hwang, and Rodrik}]{Hausmann.etal2007}
\bibinfo{author}{R.~Hausmann}, \bibinfo{author}{J.~Hwang},
  \bibinfo{author}{D.~Rodrik},
\newblock \bibinfo{title}{What you export matters},
\newblock \bibinfo{journal}{Journal of Economic Growth} \bibinfo{volume}{12}
  (\bibinfo{year}{2007}) \bibinfo{pages}{1--25}. \URLprefix
  \url{http://link.springer.com/10.1007/s10887-006-9009-4}.
  \DOIprefix\doi{10.1007/s10887-006-9009-4}.
%Type = Article
\bibitem[{Gaulier and Zignago(2010)}]{Gaulier.Zignago2010}
\bibinfo{author}{G.~Gaulier}, \bibinfo{author}{S.~Zignago},
\newblock \bibinfo{title}{{BACI}: {International} {Trade} {Database} at the
  {Product}-{Level} (the 1994-2007 {Version})},
\newblock \bibinfo{journal}{SSRN Electronic Journal}  (\bibinfo{year}{2010}).
  \URLprefix \url{http://www.ssrn.com/abstract=1994500}.
  \DOIprefix\doi{10.2139/ssrn.1994500}.
%Type = Misc
\bibitem[{{Social Progress Imperative}(2024)}]{SocialProgressImperative2024}
\bibinfo{author}{{Social Progress Imperative}}, \bibinfo{title}{Global {Social}
  {Progress} {Index} {\textbar} {Social} {Progress} {Imperative}},
  \bibinfo{year}{2024}. \URLprefix
  \url{https://www.socialprogress.org/2024-social-progress-index/}.
%Type = Misc
\bibitem[{{European Commission}(2023)}]{EuropeanCommission2023a}
\bibinfo{author}{{European Commission}}, \bibinfo{title}{Social {Progress}
  {Index}}, \bibinfo{year}{2023}. \URLprefix
  \url{https://composite-indicators.jrc.ec.europa.eu/explorer/explorer/indices/spi/social-progress-index}.
%Type = Techreport
\bibitem[{Stern et~al.(2024)Stern, Harmacek, Krylova, and
  Htitich}]{Stern.etal2024}
\bibinfo{author}{S.~Stern}, \bibinfo{author}{J.~Harmacek},
  \bibinfo{author}{P.~Krylova}, \bibinfo{author}{M.~Htitich},
  \bibinfo{title}{Social {Progress} {Index} {Methodology} {Summary}},
  \bibinfo{type}{Technical Report}, Social Progress Imperative,
  \bibinfo{address}{Washington, DC}, \bibinfo{year}{2024}. \URLprefix
  \url{https://www.socialprogress.org/static/32c1471b91c8f36250578532b3d7c802/Methodology%20Report_2024%20Social%20Progress%20Index_0.pdf}.
%Type = Techreport
\bibitem[{Block et~al.(2024)Block, Emerson, Esty, de~Sherbinin, and
  Wendling}]{Block.etal2024}
\bibinfo{author}{S.~Block}, \bibinfo{author}{J.~W. Emerson},
  \bibinfo{author}{D.~C. Esty}, \bibinfo{author}{A.~de~Sherbinin},
  \bibinfo{author}{Z.~A. Wendling}, \bibinfo{title}{Environmental {Performance}
  {Index}}, \bibinfo{type}{Technical Report}, Yale Center for Environmental Law
  \& Policy, \bibinfo{address}{New Haven, CT}, \bibinfo{year}{2024}. \URLprefix
  \url{https://epi.yale.edu/downloads/2024-epi-report.pdf}.
%Type = Misc
\bibitem[{{Yale Center for Environmental Law \&
  Policy}(2024)}]{YaleCenterforEnvironmentalLaw&Policy2024}
\bibinfo{author}{{Yale Center for Environmental Law \& Policy}},
  \bibinfo{title}{Environmental {Performance} {Index}}, \bibinfo{year}{2024}.
  \URLprefix \url{https://epi.yale.edu/}.
%Type = Misc
\bibitem[{{The World Bank}(2025)}]{TheWorldBank2025}
\bibinfo{author}{{The World Bank}}, \bibinfo{title}{World {Development}
  {Indicators}}, \bibinfo{year}{2025}. \URLprefix
  \url{https://data.worldbank.org}.
%Type = Article
\bibitem[{Jun et~al.(2020)Jun, Alshamsi, Gao, and Hidalgo}]{Jun.etal2020}
\bibinfo{author}{B.~Jun}, \bibinfo{author}{A.~Alshamsi},
  \bibinfo{author}{J.~Gao}, \bibinfo{author}{C.~A. Hidalgo},
\newblock \bibinfo{title}{Bilateral relatedness: knowledge diffusion and the
  evolution of bilateral trade},
\newblock \bibinfo{journal}{Journal of Evolutionary Economics}
  \bibinfo{volume}{30} (\bibinfo{year}{2020}) \bibinfo{pages}{247--277}.
  \URLprefix
  \url{https://ideas.repec.org//a/spr/joevec/v30y2020i2d10.1007_s00191-019-00638-7.html}.
%Type = Article
\bibitem[{Balassa(1965)}]{Balassa1965}
\bibinfo{author}{B.~Balassa},
\newblock \bibinfo{title}{Trade {Liberalisation} and "{Revealed}" {Comparative}
  {Advantage}},
\newblock \bibinfo{journal}{The Manchester School} \bibinfo{volume}{33}
  (\bibinfo{year}{1965}) \bibinfo{pages}{99--123}. \URLprefix
  \url{https://onlinelibrary.wiley.com/doi/10.1111/j.1467-9957.1965.tb00050.x}.
  \DOIprefix\doi{10.1111/j.1467-9957.1965.tb00050.x}.
%Type = Article
\bibitem[{Mealy et~al.(2019)Mealy, Farmer, and Teytelboym}]{Mealy.etal2019}
\bibinfo{author}{P.~Mealy}, \bibinfo{author}{J.~D. Farmer},
  \bibinfo{author}{A.~Teytelboym},
\newblock \bibinfo{title}{Interpreting economic complexity},
\newblock \bibinfo{journal}{Science Advances} \bibinfo{volume}{5}
  (\bibinfo{year}{2019}) \bibinfo{pages}{eaau1705}. \URLprefix
  \url{https://www.science.org/doi/10.1126/sciadv.aau1705}.
  \DOIprefix\doi{10.1126/sciadv.aau1705}.
%Type = Article
\bibitem[{Caldarelli et~al.(2012)Caldarelli, Cristelli, Gabrielli, Pietronero,
  Scala, and Tacchella}]{Caldarelli.etal2012}
\bibinfo{author}{G.~Caldarelli}, \bibinfo{author}{M.~Cristelli},
  \bibinfo{author}{A.~Gabrielli}, \bibinfo{author}{L.~Pietronero},
  \bibinfo{author}{A.~Scala}, \bibinfo{author}{A.~Tacchella},
\newblock \bibinfo{title}{A {Network} {Analysis} of {Countries}’ {Export}
  {Flows}: {Firm} {Grounds} for the {Building} {Blocks} of the {Economy}},
\newblock \bibinfo{journal}{PLoS ONE} \bibinfo{volume}{7}
  (\bibinfo{year}{2012}) \bibinfo{pages}{e47278}. \URLprefix
  \url{https://dx.plos.org/10.1371/journal.pone.0047278}.
  \DOIprefix\doi{10.1371/journal.pone.0047278}.
%Type = Article
\bibitem[{Sciarra et~al.(2020)Sciarra, Chiarotti, Ridolfi, and
  Laio}]{Sciarra.etal2020}
\bibinfo{author}{C.~Sciarra}, \bibinfo{author}{G.~Chiarotti},
  \bibinfo{author}{L.~Ridolfi}, \bibinfo{author}{F.~Laio},
\newblock \bibinfo{title}{Reconciling contrasting views on economic
  complexity},
\newblock \bibinfo{journal}{Nature Communications} \bibinfo{volume}{11}
  (\bibinfo{year}{2020}) \bibinfo{pages}{3352}. \URLprefix
  \url{https://www.nature.com/articles/s41467-020-16992-1}.
  \DOIprefix\doi{10.1038/s41467-020-16992-1}.
%Type = Misc
\bibitem[{{The World Bank}(2023)}]{TheWorldBank2023c}
\bibinfo{author}{{The World Bank}}, \bibinfo{title}{World {Bank} {Country} and
  {Lending} {Groups} – {World} {Bank} {Data} {Help} {Desk}},
  \bibinfo{year}{2023}. \URLprefix
  \url{https://datahelpdesk.worldbank.org/knowledgebase/articles/906519-world-bank-country-and-lending-groups}.
%Type = Article
\bibitem[{McFadden(1977)}]{McFadden1977}
\bibinfo{author}{D.~McFadden},
\newblock \bibinfo{title}{Quantitative {Methods} for {Analyzing} {Travel}
  {Behaviour} of {Individuals}: {Some} {Recent} {Developments}},
\newblock \bibinfo{journal}{Cowles Foundation Discussion Papers}
  (\bibinfo{year}{1977}). \URLprefix
  \url{https://elischolar.library.yale.edu/cowles-discussion-paper-series/707}.
%Type = Article
\bibitem[{Hanley and McNeil(1983)}]{Hanley.McNeil1983}
\bibinfo{author}{J.~A. Hanley}, \bibinfo{author}{B.~J. McNeil},
\newblock \bibinfo{title}{A method of comparing the areas under receiver
  operating characteristic curves derived from the same cases},
\newblock \bibinfo{journal}{Radiology} \bibinfo{volume}{148}
  (\bibinfo{year}{1983}) \bibinfo{pages}{839--843}.
  \DOIprefix\doi{10.1148/radiology.148.3.6878708}.
%Type = Article
\bibitem[{Alshamsi et~al.(2018)Alshamsi, Pinheiro, and
  Hidalgo}]{Alshamsi.etal2018}
\bibinfo{author}{A.~Alshamsi}, \bibinfo{author}{F.~L. Pinheiro},
  \bibinfo{author}{C.~A. Hidalgo},
\newblock \bibinfo{title}{Optimal diversification strategies in the networks of
  related products and of related research areas},
\newblock \bibinfo{journal}{Nature Communications} \bibinfo{volume}{9}
  (\bibinfo{year}{2018}) \bibinfo{pages}{1328}. \URLprefix
  \url{https://www.nature.com/articles/s41467-018-03740-9}.
  \DOIprefix\doi{10.1038/s41467-018-03740-9}.
%Type = Incollection
\bibitem[{Cimoli et~al.(2009)Cimoli, Dosi, and Stiglitz}]{Cimoli.etal2009}
\bibinfo{author}{M.~Cimoli}, \bibinfo{author}{G.~Dosi}, \bibinfo{author}{J.~E.
  Stiglitz},
\newblock \bibinfo{title}{The {Political} {Economy} of {Capabilities}
  {Accumulation}: {The} {Past} and {Future} of {Policies} for {Industrial}
  {Development}},
\newblock in: \bibinfo{booktitle}{Industrial {Policy} and {Development}},
  \bibinfo{edition}{1} ed., \bibinfo{publisher}{Oxford University PressOxford},
  \bibinfo{year}{2009}, pp. \bibinfo{pages}{1--16}. \URLprefix
  \url{https://academic.oup.com/book/32519/chapter/270240035}.
  \DOIprefix\doi{10.1093/acprof:oso/9780199235261.003.0001}.
%Type = Book
\bibitem[{Lee(2013)}]{Lee2013}
\bibinfo{author}{K.~Lee}, \bibinfo{title}{Schumpeterian {Analysis} of
  {Economic} {Catch}-up: {Knowledge}, {Path}-{Creation}, and the
  {Middle}-{Income} {Trap}}, \bibinfo{publisher}{Cambridge University Press},
  \bibinfo{year}{2013}.
%Type = Article
\bibitem[{Neffke et~al.(2018)Neffke, Hartog, Boschma, and
  Henning}]{Neffke.etal2018}
\bibinfo{author}{F.~Neffke}, \bibinfo{author}{M.~Hartog},
  \bibinfo{author}{R.~Boschma}, \bibinfo{author}{M.~Henning},
\newblock \bibinfo{title}{Agents of {Structural} {Change}: {The} {Role} of
  {Firms} and {Entrepreneurs} in {Regional} {Diversification}},
\newblock \bibinfo{journal}{Economic Geography} \bibinfo{volume}{94}
  (\bibinfo{year}{2018}) \bibinfo{pages}{23--48}. \URLprefix
  \url{https://doi.org/10.1080/00130095.2017.1391691}.
  \DOIprefix\doi{10.1080/00130095.2017.1391691}.
%Type = Article
\bibitem[{Castaldi et~al.(2015)Castaldi, Frenken, and Los}]{Castaldi.etal2015}
\bibinfo{author}{C.~Castaldi}, \bibinfo{author}{K.~Frenken},
  \bibinfo{author}{B.~Los},
\newblock \bibinfo{title}{Related {Variety}, {Unrelated} {Variety} and
  {Technological} {Breakthroughs}: {An} analysis of {US} {State}-{Level}
  {Patenting}},
\newblock \bibinfo{journal}{Regional Studies} \bibinfo{volume}{49}
  (\bibinfo{year}{2015}) \bibinfo{pages}{767--781}. \URLprefix
  \url{https://doi.org/10.1080/00343404.2014.940305}.
  \DOIprefix\doi{10.1080/00343404.2014.940305}.

\end{thebibliography}




% %% For numbered reference style

%   \textit{\LaTeX: a document preparation system},
%   Addison Wesley, Massachusetts,

% \end{thebibliography}
\end{document}